\documentclass[double]{article}
\usepackage{times}
\usepackage{algorithm}
\usepackage{algorithmic}
\usepackage{wrapfig}
\usepackage{verbatim}
\usepackage{amsmath}
\usepackage{graphicx}
\usepackage{subcaption}
\usepackage[active]{srcltx}
\usepackage{color}
\usepackage{lineno}
\usepackage{dirtytalk}
\usepackage{amsthm}
\usepackage{authblk}
\usepackage{listings}
\usepackage{tikz}
\usepackage{longtable}
\usetikzlibrary{shapes.geometric, arrows}

\newtheorem{definition}{Definition}

\usepackage{epsfig,graphicx,amsfonts}
\usepackage{amsmath,amsthm,amssymb}
\usepackage{xcolor}

\usepackage{adjustbox}
\usepackage{multirow}
\usepackage{setspace}
\usepackage{hyperref}
\usepackage{comment}

\newcommand{\review}[1]{\textcolor{black}{#1}}

\long\def\comment #1\commentend{}

\begin{document}

\title{\Large Service Preservation from Matching Non-Matching Socks Under Stochastic Loss}

\author{Teddy Lazebnik$^{1,2,*}$\\ \(^1\) Department of Information Science,  University of Haifa, Haifa, Israel\\ \(^2\) Department of Computing, Jonkoping University, Jonkoping, Sweden \\ \(^*\) Corresponding author: teddy.lazebnik@ju.se \\ }

\date{ }

\maketitle 

\begin{abstract}
\noindent
\review{Socks are produced and replaced at a massive scale, yet their paired use makes them unusually vulnerable to service loss, as the disappearance of a single sock can leave usable wear-capacity stranded and create sockless days even when functional socks remain available. In this study, we examine whether pairing non-matching \say{orphan} socks can preserve daily sock service under stochastic loss, and how this benefit trades off against perceived social discomfort. We formalize sock ownership as a sequential decision problem under uncertainty in which socks wear out and disappear stochastically during laundering, while public exposure induces a person-specific mismatch penalty. We conducted an in-person study to estimate mismatch sensitivity and diversity preference, linking behavioural heterogeneity to interpretable mixing policies. Using these results, an exact benchmark on small tractable instances, and a simulation-based evaluation of pairing policies, we show that strict matching can appear resource-frugal largely because it generates many sockless days. In contrast, controlled tolerance for mismatch sustains service and reduces stranded wear-capacity across loss regimes. The ecological-cost term is treated as a proportional embodied-cost proxy rather than an independent life-cycle assessment measure, so the environmental interpretation is suggestive and mechanism-based rather than a direct estimate of environmental savings. This study establishes the feasibility and limitations of matching non-matching socks as a simple strategy for preserving service from already-owned garments.} \\ \\

\end{abstract}

\maketitle \thispagestyle{empty}
\pagestyle{myheadings} \markboth{Draft:  \today}{Draft:  \today}
\setcounter{page}{1}

\section{Introduction}
\label{sec:introduction}
Socks first appear in historical records at least as early as the \(8^{\text{th}}\) century, although it is widely believed that they were worn earlier than the surviving written evidence suggests \cite{bush1994folk}. Socks are a type of clothing worn on the feet designed to provide warmth, comfort, and protection to the feet and lower legs \cite{intro_1,intro_2}. They come in various styles, lengths, and materials, and are worn for both practical and aesthetic purposes \cite{intro_3}. They are a fundamental component of everyday life as people all over the world wear them on a daily basis \cite{intro_4,intro_5}. Despite their humble reputation as \say{background clothing}, socks constitute a substantial consumer market: for 2024, the global socks market is estimated at 14.66 billion USD, with roughly 3.8 billion pairs sold\footnote{\url{https://www.statista.com/outlook/cmo/apparel/men-s-apparel/socks/worldwide}}. 

While the environmental discussion around apparel often focuses on conspicuous items (e.g., jeans, shoes, coats), the textiles sector's footprint is largely driven by scale with high volumes of production, frequent replacement, and short effective lifetimes of garments \cite{Niinimaki2020FastFashion}. This footprint is not limited to manufacturing; consumer use (washing, drying, ironing) also contributes materially to energy and water demand and pollution \cite{Sajn2022TextilesEnvironment}. Consequently, major sustainability road-maps emphasize keeping textiles in use for longer and extracting more value per manufactured item - a central principle of circular-economy thinking in fashion \cite{nguyen2025global}. Global policy and advocacy efforts similarly call for interventions that reduce needless replacement and waste across the apparel lifecycle.

Unlike other clothing, socks have a unique attribute as they are worn in pairs and there are multiple pairs in one's possession, unlike other paired cloths, such as gloves, that are usually more scarce \cite{intro_6}. Moreover, while not officially established in the literature, additional common phenomena regarding socks are their unexpected yet often disappearing nature. In particular, laundry is a common setting where socks tend to disappear, more often than not leaving a single sock without its original pair \cite{socks_dis_2}. As such, socks provide a surprisingly clean microcosm of this macro problem. Indeed, in everyday laundering and household routines, socks may disappear or become separated, leaving behind functional \say{orphan} socks without their original mates. Regardless of whether socks are lost to laundry mechanics, household entropy, or gremlins, the practical outcome is the same: ownership drifts from paired sets toward a growing pile of single socks.

This creates a decision dilemma with economic, ecological, and social dimensions. Economically and environmentally, discarding a usable orphan sock (or replacing a perfectly functional pair because one sock vanished) is wasteful; socially, wearing intentionally non-matching socks can be interpreted as a minor norm violation \cite{intro_7,intro_8}. Classic sociological accounts of impression management frame clothing as a cue used to signal conformity, competence, and group membership in everyday interactions \cite{Goffman1959Presentation}. Complementing this view, experimental work in social psychology demonstrates that clothing can carry symbolic meaning with measurable downstream effects on perception and behavior \cite{AdamGalinsky2012Enclothed}. In short, mismatching can be ecologically sensible while being socially \say{expensive}. Nonetheless, this strategy can be remedied, to some degree, by performing a better match between different socks. For example, let us consider the naive scenario when a sock owner owns 10 identical socks (from 5 pairs). In such an example, pairing any two socks from the set will result in zero social price. However, this case does not align with the usage of cloth, in general, and socks, in particular, for self-expression \cite{intro_9,intro_10}. 

\review{In this study, we ask a deliberately practical question: can pairing non-matching socks preserve daily sock service under stochastic loss, and at what perceived social cost? In order to address this question, we propose a computational model in which an agent (a sock owner) chooses an initial sock portfolio and then repeatedly selects daily pairs under uncertainty due to wear-out and stochastic disappearance. The resulting decision problem fits the optimization under uncertainty setting \cite{ouu_review}; however, our contribution is not a closed-form derivation of an optimal policy for the full stochastic problem. Instead, we use the formal model to define the objective, state dynamics, and constraints, prove computational hardness for the general problem, solve small tractable instances exactly as a benchmark, and evaluate scalable interpretable policies through simulation. We complement the model with behavioral data: (i) pairwise comparisons of sock pairs are used to estimate individual mismatch sensitivity, interpreted as anticipated social discomfort, and (ii) bundle choices are used to estimate preferences over variety versus matchability. Our field component further tests whether participants will wear visibly mismatched socks in public to evaluate the feasibility of the proposed behavior within the sampled context. Monetary and ecological quantities are reported as secondary proxy outcomes, while the central empirical outcome is service preservation, measured through infeasible days and stranded wear-capacity.}

The rest of this paper is organized as follows. Section~\ref{sec:related} surveys relevant work on sock markets, textiles’ ecological impacts, the social interpretation of sartorial nonconformity, and optimization under uncertainty. Section~\ref{sec:model} formally presents the proposed model. \review{Section~\ref{sec:exp} describes our evaluation approach, combining exact optimization on small tractable instances, simulations over a range of loss and wear regimes, and an in-person user study to estimate key behavioral parameters.} Section~\ref{sec:results} presents the empirical and simulation results. Finally, Section~\ref{sec:discussion} discusses implications, limitations, and practical takeaways.

\section{Related Work}
\label{sec:related}
The global hosiery and sock industry represents a profound intersection of macroeconomic supply chain dynamics and nuanced socio-psychological consumer behaviors. Traditionally relegated to a low-involvement commodity sector, the production and utilization of socks have evolved into a highly specialized global enterprise. 

\subsection{Macroeconomic valuation and market dynamics of the sock industry}
The economic trajectory of the global sock market is characterized by robust, compounding growth, driven by a confluence of evolving consumer aesthetics, the premiumization of functional apparel, and the rapid expansion of digital retail architectures. Market valuation models uniformly project significant expansion over the coming decade, underscoring the transition of hosiery from a basic utilitarian necessity to a lifestyle-enhancing accessory \cite{fortune2025socks}.

Empirical market analyses establish a baseline valuation for the global socks market in 2024 and 2025 at approximately USD 47.1 billion to USD 56.7 billion \cite{grandview2025socks}. The projected growth over the forecast period extending to 2033–2036, illustrates an aggressive upward trajectory. Moderate predictive models estimate the market will reach USD 73.83 billion by 2033, maintaining a Compound Annual Growth Rate (CAGR) of 5.2\% \cite{grandview2025socks}. Conversely, more bullish economic forecasts anticipate valuations ranging from USD 98.27 billion to USD 109.6 billion by 2033 and 2034, driven by a CAGR of approximately 6.9\% to 7.0\% \cite{fortune2025socks}. The expansion is structurally supported by the pervasive adoption of hybrid work models, which have cemented the dominance of the casual and lifestyle sock segments, currently accounting for over 60\% of global revenue shares \cite{gmi2025socks}. 

Simultaneously, the athletic and performance segment is poised to exhibit the most rapid acceleration. Driven by global fitness cultures and the athleisure aesthetic, this segment benefits from advanced textile engineering, including moisture-wicking polymers, temperature-regulating smart textiles, and graduated compression features. The integration of antimicrobial treatments, such as copper-infused materials, further caters to health-conscious and aging demographics requiring therapeutic diabetic hosiery \cite{fmi2025socks}.

Geographically, the Asia-Pacific region dominates the global landscape, holding the largest revenue share, historically ranging from 37.2\% to 38.46\% and exhibiting the fastest regional growth \cite{grandview2025socks}. This dominance is intrinsically linked to rising middle-class disposable incomes, rapid urbanization, and an entrenched manufacturing infrastructure in nations such as China and India. Europe and North America represent mature markets driven by premiumization; Europe excels in the adoption of sustainable, ethically sourced textiles, while North American consumers demonstrate a high willingness to pay premiums for durability, compression support, and shape retention \cite{fortune2025socks, pmr2025socks}.

\subsection{The social cost of non-matched socks}
Despite the fact that macroeconomic value chains and ecological metrics define the production of hosiery, the terminal phase of the sock's life cycle occurs within the domestic sphere. Here, the sock becomes the focal point of a pervasive sociological and psychological phenomenon: the attrition of the single sock, the resultant cognitive labor required to manage garment inventory, and the social costs associated with wearing non-matched items. Statistical modeling indicates that the average consumer loses approximately 1.3 socks per month, culminating in a lifetime financial cost of thousands of dollars and resulting in millions of perfectly viable, orphaned socks being prematurely discarded into landfills. However, the adoption of mismatched hosiery is heavily regulated by psychological principles of social cognition, impression management, and professional stigma. The societal reception of mismatched garments hinges entirely on the perception of intentionality.

When conformity to sartorial norms is expected, individuals dress to gain social acceptance and signal competence. According to stigma frameworks, failing to meet these aesthetic standards incurs a significant \say{social cost} \cite{goffman1963stigma}. If an observer perceives the wearing of mismatched socks as an unintentional error, the social consequences are overwhelmingly negative. An unintentional mismatch signals a failure of executive function, a lack of resources, or a deficit in bounded rationality \cite{lsu2025stigma}. Evaluators frequently equate an inability to manage basic personal presentation with an inability to manage professional responsibilities, resulting in severe penalties regarding perceived competence and reliability \cite{gurung2018dressing}.

This dynamic also operates internally via the psychological phenomenon of \say{enclothed cognition}. Research demonstrates that the physical experience of wearing specific garments profoundly influences a subject's psychological state and cognitive processing style \cite{slepian2015enclothed}. Wearing formal, conforming clothing enhances abstract cognitive processing and fosters feelings of professional distance and power \cite{slepian2015enclothed}. Conversely, an individual realizing they are unintentionally wearing mismatched garments experiences self-objectification and acute anxiety, which can severely disrupt cognitive focus and impair performance in professional environments \cite{jinnah2025fashion}.

In stark contrast to the penalties of unintentional untidiness, the deliberate and intentional flouting of sartorial norms yields an inverted psychological result. This paradigm is formally established in behavioral economics as the \say{Red Sneakers Effect} \cite{bellezza2014red}. The Red Sneakers Effect posits that under specific boundary conditions, visible and costly acts of nonconforming behavior—such as an executive deliberately wearing wildly patterned, mismatched socks with a bespoke suit—lead observers to infer higher status and competence than if the individual had strictly conformed \cite{bellezza2014red}. Anchored in evolutionary signaling theory and the economics of conspicuous consumption, this phenomenon occurs because nonconformity carries an inherent social risk of alienation. Consequently, observers infer that an individual who voluntarily embraces this risk must possess an abundance of social capital, power, and autonomy. The mismatched socks act as a costly signal, demonstrating that the wearer’s hierarchical position is so secure that they do not need to rely on the safety of conformity \cite{bellezza2014red}.

The efficacy of this status signaling, however, is governed by strict parameters. The nonconformity must be overtly deliberate; if the observer suspects the mismatch is a mere accident of the laundry cycle, the positive inferences of status instantly collapse into negative judgments of incompetence \cite{bellezza2014red}. Furthermore, the effect is moderated by the observer's own \say{need for uniqueness} and relies heavily on the existence of a strict baseline of formal conduct against which the nonconformity can contrast \cite{bellezza2014red}. By transforming a logistical household failure into a strategic asset of personal branding, the intentional wearing of mismatched socks circumvents both the gendered cognitive labor of laundry sorting and the traditional social costs of untidiness, effectively converting vulnerability into an emblem of autonomous power.

\subsection{Optimization under uncertainty}
The socks buying and matching task can be formalized as an optimization under uncertainty (OUU), an active field of research in computer science \cite{ouu_review_2}. Many real-world problems take the form of OUU tasks due to one or more unknown or stochastic elements in the problem \cite{ouu_g_1,ouu_g_2,ouu_g_3}. For example, \cite{ouu_supply_chain_example} proposed a scheme based on a two-stage stochastic program formulation that addresses the optimization of an energy supply chain in the presence of uncertainties. \cite{ouu_hospital} presented an agent-based simulation with a deep reinforcement learning agent for the hospital staff and resources allocation task, where the patient's flow and disease distribution is unknown in advance and stochastic. \cite{ouu_disaster} formalize an OUU task for the capacitated vehicle routing problem with uncertain travel times and demands when planning vehicle routes for delivering critical supplies to a population in need after a disaster. 

A typical OUU formulation constitutes a double-loop algorithm, wherein, an uncertainty quantification \cite{ouu_define} procedure is nested inside an outer optimization loop. In OUU, two types of variables are considered - design variables and random variables. A generalized OUU formulation takes the form:
\begin{equation}
    \max_{x \in X} f(x, \alpha) \; \text{ s.t. } P[\{g_1(x, \alpha) \leq 0, \dots, g_n(x, \alpha) \leq 0\}] \leq P_{f_t},
    \label{eq:ouu_background}
\end{equation}
where \(x\) is the vector of design variables, \(\alpha\) is the vector of random variables, \(f(x, \alpha)\) is the objective function, \(g_i\) are the constraints (both probabilistic and deterministic), \(P[\cdot]\) is the probability operator, \(P_{f_t}\) is the admissible/target failure probability, and \(N\) is the number of constraints. 

One of the leading approaches to tackling OUU is the sampling-based approach. It draws its popularity from its simple implementation while obtaining comparable performance for much more sophisticated solutions \cite{ouu_methods}. Sampling-based approaches predominantly include Monte Carlo simulation (MCS) techniques \cite{ouu_mc} or its variants, such as separable Monte Carlo \cite{ouu_exp_1} and the multilevel Monte Carlo approaches \cite{ouu_exp_2}. If the probability information for the inputs is known, MCS is applied directly in its basic form when the analysis model can be evaluated with low computational cost, such as the one proposed in this study. 

\section{Model Definition}
\label{sec:model}
We formalize sock ownership and daily pairing as a sequential decision-making OUU problem, where socks may become unusable due to wear-out and may also disappear stochastically during laundering. The goal is to quantify the trade-off between (i) economical-ecological savings achieved by allowing non-matching pairs and (ii) the social cost incurred by visibly mismatching socks in public.

Mathematically, a \textit{sock} is defined by a tuple \(s := (p, e, \tau, \theta, d, \overrightarrow{u})\), where \(p \in \mathbb{R}^+\) is the monetary price, \(e \in \mathbb{R}^+\) is an embodied ecological cost proxy (e.g., CO\(_2\)e units per sock; used as a weight in the objective), \(\tau \in \mathbb{N}\) is the number of times the sock has been worn so far, \(\theta \in \mathbb{N}\) is the maximum number of wears before the sock is deemed unwearable, \(d \in [0,1]\) is the probability that the sock disappears when washed, and \(\overrightarrow{u}\in\mathcal{U}\) is a feature vector describing appearance (e.g., color, pattern, length, material), where \(\mathcal{U}\) is a finite discrete product space. To this end, let \(S(t)\) be the set (multiset) of socks available to the agent at time \(t\), and let \(L(t)\) be the multiset of socks currently in the laundry buffer.

For this model, we distinguish between \textit{appearance mismatch} and \textit{social penalty}. Namely, appearance mismatch depends on the two socks, while social penalty depends on the person and on whether the situation is public. Hence, we define an appearance dissimilarity function \(\eta(s_i,s_j) \in [0,1]\), where \(\eta=0\) indicates identical-looking socks (maximally matching), and \(\eta=1\) indicates maximally non-matching socks. For convenience, we also define a compatibility score \(\xi(s_i,s_j) := 1 - \eta(s_i,s_j) \in [0,1]\).  Let \(Z_t \in \{0,1\}\) be some public exposure indicator at time \(t\), where \(Z_t=1\) means the agent expects the socks to be publicly observable (and thus socially relevant). We assume \(Z_t \sim \text{Bernoulli}(\rho)\) i.i.d., where \(\rho\in[0,1]\) is the agent's exposure rate.

Each agent has a nonnegative \say{mismatch sensitivity} parameter \(\chi \in \mathbb{R}^+\) capturing the strength of the social penalty they experience when wearing mismatched socks in public. The per-day social cost for wearing socks \((s_i,s_j)\) at time \(t\) is
\begin{equation}
C_{\text{soc}}(s_i,s_j; \chi, Z_t) \;=\; Z_t \cdot \chi \cdot g(\eta(s_i,s_j)),
\label{eq:social_cost}
\end{equation}
where \(g:[0,1]\rightarrow\mathbb{R}^+\) is a monotone increasing penalty shape (e.g., \(g(x)=x\) linear, \(g(x)=x^2\) convex, or a threshold rule).

Furthermore, in order to capture the notion that some agents actively prefer variety in their sock collection (even if it reduces matchability), each agent has a \emph{diversity preference} parameter \(\delta \in \mathbb{R}^+\). Thus, let \(D(S)\) be a diversity functional over a sock multiset. We use a generic definition here, and instantiate it in the experiments/simulations section (e.g., Shannon diversity over discrete appearance categories, or mean pairwise feature dispersion). The diversity term contributes a benefit (utility) that grows with diversity:
\begin{equation}
U_{\text{div}}(S;\delta) \;=\; \delta \cdot D(S).
\label{eq:diversity_utility}
\end{equation}

Based on this dynamics, an agent is defined by the tuple \(a := (b, \chi, \delta, \rho)\), where \(b\in\mathbb{R}^+\) is the available monetary budget (or initial budget, depending on the buying policy), \(\chi\) is mismatch sensitivity, \(\delta\) is diversity preference, and \(\rho\) is exposure rate. At each day \(t\in\{1,\dots,T\}\), the agent selects an \emph{action} consisting of \(A_t := \big( s_t^1, s_t^2, B_t \big)\), where \(s_t^1,s_t^2 \in S(t)\) are the two socks chosen to wear (distinct physical socks), and \(B_t \subseteq \mathbb{S}\) is an optional set of socks purchased at time \(t\) from an available catalogue \(\mathbb{S}\) (possibly empty). Purchases incur monetary and ecological costs. 

We define the per-day immediate utility as
\begin{equation}
R_t \;=\; -\sum_{s\in B_t}\big(p(s) + \lambda \, e(s)\big)\;-\; C_{\text{soc}}(s_t^1,s_t^2;\chi,Z_t),
\label{eq:reward}
\end{equation}
where \(\lambda \ge 0\) is a scalar converting ecological cost units into the utility scale. Optionally, diversity can be incorporated as an episodic utility term (e.g., evaluated at \(t=0\) for the initial collection, or periodically):
\[
R_t \leftarrow R_t + \mathbb{I}[t\in\mathcal{T}_{\text{div}}]\cdot U_{\text{div}}(S(t);\delta),
\]
for some evaluation times \(\mathcal{T}_{\text{div}}\) (commonly \(\{0\}\) or \(\{T\}\)).

The agent's goal is to maximize expected total utility over horizon \(T\):
\begin{equation}
\max_{\pi}\; \mathbb{E}\left[\sum_{t=1}^{T} R_t \right]
\quad \text{s.t.}\quad
\sum_{t=1}^{T}\sum_{s\in B_t} p(s) \le b,
\label{eq:objective}
\end{equation}
where \(\pi\) is a policy mapping the observable state to an action. \review{Specifically, Eq.~(\ref{eq:objective}) defines the normative optimization problem used as a reference point. In the full model, the state space grows combinatorially with the number of socks, usage counters, laundry-buffer configurations, and remaining budget. Therefore, we do not claim to derive a closed-form optimal policy for the general case. Instead, we use this objective in two complementary ways: first, as an exact benchmark on deliberately small finite-horizon instances, and second, as the criterion motivating the scalable heuristic policies evaluated by Monte Carlo simulation.}

Based on this formalization, the system state at time \(t\) takes the form: \(
X(t) := \big(S(t), L(t), b(t)\big)\), where \(b(t)\) is remaining budget (if purchases are allowed after \(t=0\)). When the agent wears \((s_t^1,s_t^2)\), both socks' usage counters increase: \(\tau(s_t^k) \leftarrow \tau(s_t^k)+1,\quad k\in\{1,2\}\). If \(\tau(s)>\theta(s)\), the sock is removed permanently (wear-out). The worn socks are moved from \(S(t)\) into the laundry buffer \(L(t)\).

Laundry is processed when the buffer reaches capacity \(\kappa\in\mathbb{N}\) (a model parameter): \(\text{if } |L(t)| \ge \kappa \text{ then wash } L(t)\).  Upon washing, each sock \(s\in L(t)\) returns to \(S(t+1)\) independently with probability \(1-d(s)\); otherwise it disappears with probability \(d(s)\). After processing, the laundry buffer is cleared (or reduced, depending on whether more than \(\kappa\) socks were queued).

In addition to the utility in Eq.~(\ref{eq:objective}), we report interpretable, directly measurable outcomes from simulations and experiments, including: (i) total monetary spend \(\sum p(s)\), (ii) total embodied ecological cost \(\sum e(s)\), (iii) total wears achieved (proxy for service delivered), and (iv) \say{stranded wear-capacity}—the unused remaining \(\theta(s)-\tau(s)\) of socks that are discarded or rendered effectively unusable due to pairing constraints. These measures let us quantify the economical--ecological benefit of tolerating non-matching socks, and relate it to the social cost through \(\chi\) and exposure \(\rho\).

Proofs that the proposed model is NP-hard as well as hard to approximate are provided in the Appendix.

\section{Experimental Design}
\label{sec:exp}
\review{We adopted a combined optimization, simulation, and behavioral evaluation approach. First, we solve small finite-horizon instances exactly in order to benchmark the proposed heuristic policies against an optimal policy whenever the state space is tractable. Second, we conduct a larger \textit{in silico} simulation study that evaluates interpretable matching strategies under multiple loss, wear, exposure, and preference configurations. Third, we conducted a user experiment which serves two purposes: it provides direct evidence for the usability of the proposed method and generates the data needed to obtain realistic behavioral parameters for the simulations.}

\subsection{Simulation Study}
\label{sec:sim}
\review{Since the optimization problem induced by our model is NP-hard and even hard to approximate, exact optimization is not computationally practical for the full-scale catalogue and horizon used in the main experiments. Nevertheless, to address the gap between a formal optimization model and purely heuristic simulation, we include an exact benchmark on small tractable instances. These instances are solved by finite-horizon dynamic programming and are used to estimate the optimality gap of the proposed policies. For the full-scale setting, we then evaluate practical decision policies via Monte Carlo simulation \cite{harrison2010introduction}. The simulation is designed to quantify the economical-ecological benefit of allowing non-matching socks, and relate it to the incurred social cost, across a wide range of disappearance and wear regimes.}

\subsubsection{Simulation environment and initialization}
\label{sec:sim_env}
For the simulation, \(\tau\) is initialized to \(0\) upon purchase. When not using empirical distributions, we generate catalogue instances by sampling:
(i) discrete appearance features \(\overrightarrow{u}\) from a finite product space \(\mathcal{U}\), (ii) prices \(p\) from a bounded integer range, (iii) wear limits \(\theta\) from a chosen positive integer distribution, and (iv) disappearance probabilities \(d\) from \([0,1]\). \review{Embodied ecological cost \(e\) is modeled as a nonnegative scalar proxy rather than as a full life-cycle assessment. Unless stated otherwise, we set \(e(s)=\alpha\cdot p(s)\) for a fixed \(\alpha>0\), using price as a neutral embodied-cost proxy in the absence of product-specific life-cycle assessment data. This proxy is used to study how pairing policies affect replacement pressure and stranded usable capacity; it does not estimate product-specific emissions, water use, textile composition, transport, washing and drying behavior, or disposal pathways. We also report results in units of \say{socks purchased} and stranded wear-capacity to avoid dependence on any particular footprint coefficient. }

\review{Notably, the ecological term \(C_{\text{eco}}\) should be interpreted as an embodied-cost proxy rather than as a complete environmental footprint. In the simulator, \(e(s)\) is a scalar weight attached to purchasing or replacing a sock, and is used to quantify replacement pressure and unused wear-capacity under different pairing policies. The model therefore captures the mechanism by which tolerating mismatch can extend the use of already-owned socks, but it does not claim to measure realized emissions, water use, or landfill diversion.}

Each simulation instance is parameterized by an agent tuple \(a=(b,\chi,\delta,\rho)\) (Section~\ref{sec:model}), a horizon \(T\), a laundry capacity \(\kappa\), and a policy \(\pi\) (defined below). The initial state is
\[
X(0)=(S(0),L(0),b(0)),
\]
where \(L(0)=\emptyset\), \(b(0)=b\), and \(S(0)\) is determined by the policy's purchase rule \(\pi_{\text{buy}}\).

Each day \(t\in\{1,\dots,T\}\), the simulator performs five steps. First, draw exposure \(Z_t \sim \text{Bernoulli}(\rho)\). Second, select a pair \((s_t^1,s_t^2)\) using \(\pi_{\text{pair}}\), possibly after purchasing \(B_t\) if replenishment is enabled. Third, apply wear update following \(\tau(s_t^k)\leftarrow \tau(s_t^k)+1\) for \(k\in\{1,2\}\) and remove any \(s\) with \(\tau(s)>\theta(s)\). Fourth, move worn socks to the laundry buffer \(L(t)\). Fifth, if \(|L(t)|\ge \kappa\), wash: each \(s\in L(t)\) returns to \(S(t+1)\) independently with probability \(1-d(s)\), otherwise it disappears; then clear \(L(t)\). If at any day \(|S(t)|<2\) and replenishment is not allowed (or the budget is exhausted), the day is marked as \emph{infeasible} (no socks to wear), and no wear/laundry updates are applied.

\subsubsection{Policies - matching and purchasing strategies}
\label{sec:sim_policies}
A policy \(\pi\) is composed of two components: \(\pi = (\pi_{\text{buy}}, \pi_{\text{pair}})\), where \(\pi_{\text{buy}}\) selects socks to purchase (either only at \(t=0\), or also as replenishment when allowed), and \(\pi_{\text{pair}}\) selects a daily pair \((s_t^1,s_t^2)\) from the available inventory \(S(t)\). 

We evaluate the following families of strategies, all of which are computationally light and interpretable:

\paragraph{(P1) Purist (strict matching).}
The agent is unwilling to wear visibly non-matching socks. Formally, it only permits pairs whose mismatch satisfies \(\eta(s_i,s_j)\le \tau_{\eta}\), where \(\tau_{\eta}\) is a strict threshold (often \(\tau_{\eta}=0\) for \say{identical only}). If no permissible pair exists, the agent (i) buys socks if purchases are allowed, otherwise (ii) incurs a \say{no-pair} day (defined below).

\paragraph{(P2) Greedy-Compatibility.}
Each day, select the pair with maximal compatibility:
\[
(s_t^1,s_t^2)=\arg\max_{x\neq y \in S(t)} \xi(x,y).
\]
This policy ignores future dynamics and social exposure stochasticity, serving as a simple baseline for ``best-looking pair today''.

\paragraph{(P3) Threshold-Mix.}
If a highly compatible pair exists (e.g., \(\xi\ge \tau_{\xi}\)), wear it. Otherwise, select the best available pair even if mismatched:
\[
(s_t^1,s_t^2)=\arg\max_{x\neq y \in S(t)} \xi(x,y)
\quad\text{subject to}\quad
\xi(x,y)\ge\tau_{\xi}\ \text{if feasible}.
\]
This policy interpolates between purist behavior (high \(\tau_{\xi}\)) and greedy mixing (low \(\tau_{\xi}\)).

\paragraph{(P4) Orphan-Rescue.}
Prioritize consuming orphan socks (socks with few compatible partners remaining) to reduce stranded wear-capacity. Let
\[
\text{deg}_\tau(s) := \left|\{\, s'\in S(t)\setminus\{s\} \;:\; \xi(s,s')\ge\tau_{\xi}\,\}\right|
\]
be the number of acceptable partners under a threshold \(\tau_{\xi}\).
The policy first selects a sock \(s\) with minimal \(\text{deg}_\tau(s)\) (``most at risk of becoming stranded''), then pairs it with the partner \(s'\) maximizing \(\xi(s,s')\).

\paragraph{(P5) Exposure-Aware Greedy.}
When public exposure is likely (large \(\rho\) and/or large \(\chi\)), mismatch should be avoided. This policy selects a pair that greedily maximizes an estimate of immediate utility:
\[
(s_t^1,s_t^2)=\arg\max_{x\neq y\in S(t)}\left[\ \xi(x,y)\;-\;\hat{Z}_t\cdot \chi \cdot g(\eta(x,y))\ \right],
\]
where \(\hat{Z}_t:=\rho\) is the expected exposure indicator (since \(Z_t\sim \text{Bernoulli}(\rho)\)).

In this context, we consider two purchase rules: (i) initial-only purchase where all socks are purchased at \(t=0\), and
(ii) replenishment allowed where the agent may purchase additional socks \(B_t\) as long as total spend remains \(\le b\). For replenishment policies, we use simple rules such as when the number of available socks \(|S(t)|\) drops below \(2\), buy the cheapest two socks that maximize \(\xi\) (or that minimize expected social cost under \(\rho,\chi\)) or buy socks from a restricted subset of features (low expected \(\eta\) to many existing socks) to increase future matchability.

\subsubsection{Exact benchmark on tractable instances}
\review{To ensure that the proposed policies are not evaluated only as unbenchmarked heuristics, we additionally solve small instances of the model exactly. The exact benchmark uses the same state variables and immediate reward, but restricts the catalogue size, horizon, and budget so that the full finite-horizon state space can be enumerated. Specifically, a state is represented as
\(X(t)=(S(t),L(t),b(t))\), where (\(S(t)\)) is the available sock multiset, (\(L(t)\)) is the laundry buffer, and (\(b(t)\)) is the remaining budget. For each state, the feasible action set contains all possible daily pairs (\(s_i,s_j\)) with \(s_i\neq s_j\), together with any admissible purchase action \(B_t\) that respects the remaining budget. The optimal value function is computed by backward induction:
\[
V_t(X)=\max_{A\in\mathcal{A}(X)}
\mathbb{E}\left[
R_t(X,A,Z_t)+V_{t+1}(X')
\mid X,A
\right],\]
with terminal condition \(V_{T+1}(X)=0\). The expectation is taken over public exposure \(Z_t\) and laundry disappearance events. When the laundry buffer is washed, all return/loss outcomes are enumerated exactly for the small benchmark instances.}

\review{Because exact enumeration becomes infeasible as the number of socks and the horizon grow, the benchmark is intentionally limited to small instances. These instances are not intended to replace the full simulation study. Rather, they serve as a diagnostic benchmark that reports how far each heuristic is from the true optimum when the optimum is computable. For each benchmark instance and policy \(\pi\), we report the relative optimality gap \(\mathrm{Gap}(\pi)= (V^{*}-V^{\pi})/({|V^{*}|+\epsilon})\), 
where (\(V^{*}\)) is the exact optimal value, (\(V^{\pi}\)) is the value obtained by policy (\(\pi\)), and (\(\epsilon>0\)) is a small numerical stabilizer.}

\subsubsection{Statistical analysis}
\label{sec:sim_sweeps}
We fix a horizon \(T\) and a catalogue generation scheme for \(\mathbb{S}\) (or use an empirical catalogue when available). Each purchased sock has \(\tau=0\), disappearance probability \(d\) (either homogeneous or type-dependent), and wear-limit \(\theta\). Public exposure is drawn i.i.d.\ as \(Z_t\sim \text{Bernoulli}(\rho)\).
For policies that require a threshold (e.g., Threshold-Mix), we use \(\tau_{\xi}\) as a control knob, varied explicitly when constructing trade-off curves. In this context, we conduct \review{four} experiments to evaluate the proposed model. \review{The first experiment is the exact tractable-instance benchmark described in Section~\ref{sec}. For each small instance, we compute the optimal finite-horizon policy and compare Purist, Greedy-Compatibility, Threshold-Mix, Orphan-Rescue, and Exposure-Aware Greedy against this optimum using the optimality gap and the interpretable outcome measures.} For the \review{second} experiment, we fix one reference configuration \(\Theta_{\text{ref}}\) (catalogue size, \(T\), \(\kappa\), \(b\), \(\theta\), \(d\), \(\rho\), \(\chi\), \(\delta\)). We report, per policy, the mean and CI of socks purchased (and \(C_{\$}\)), ecological proxy \(C_{\text{eco}}\), social cost \(C_{\text{soc}}\), infeasible days, and stranded wear-capacity \(C_{\text{strand}}\). This provides a single comparison of policies under a mid-range \say{typical} scenario. For the \review{third} experiment, we vary \(d \in \mathcal{D} \quad\text{and}\quad \theta \in \Theta_{\text{wear}}\), holding other parameters fixed (including \(\rho,\chi,\delta\)). For each \((d,\theta)\), we run all policies and plot expected socks purchased vs.\ \(d\), expected infeasible days vs.\ \(d\), and stranded capacity vs.\ \(d\). We repeat for each wear regime \(\theta\in\Theta_{\text{wear}}\). This allows to quantify how the benefit of mixing grows as laundry loss increases and as socks wear out faster. Lastly, for the \review{forth} experiment, for each policy family with a tunable tolerance (e.g., Threshold-Mix, Orphan-Rescue with a threshold), we sweep: \(\tau_{\xi} \in \mathcal{T}_{\xi}\), and compute the trade-off points:
\[
(\Delta C_{\text{soc}}(\tau_{\xi}),\ \Delta C_{\$}(\tau_{\xi})) \quad\text{and}\quad
(\Delta C_{\text{soc}}(\tau_{\xi}),\ \Delta C_{\text{eco}}(\tau_{\xi})),
\]
relative to the Purist as baseline. We report the resulting curves and identify: (i) the knee-point (largest marginal savings per unit social cost), and (ii) the Pareto-dominant region. This explicitly indicates the trade-off between ecological/economic benefit and the associated social cost. A description of the simulation initialization with real-world data is provided in the Appendix.

\subsection{User Experiments}
\label{sec:exp_user}
The user experimental component of this study has two objectives. First, we estimate each participant's mismatch sensitivity (\(\chi\)), which operationalizes the social cost incurred when wearing visually non-matching socks in public (Eq.~(\ref{eq:social_cost})). Second, we estimate each participant's diversity preference (\(\delta\)), which captures the extent to which they prefer varied sock collections even at the expense of reduced matchability (Eq.~(\ref{eq:diversity_utility})). To this end, we conduct (i) an incentivized pairwise-comparison task over sock pairs (to learn \(\chi\)) and (ii) an incentivized bundle-preference task over sock sets with controlled diversity (to learn \(\delta\)).

\subsubsection{Participants}
Participants were recruited via social media announcements (e.g., community groups and institutional channels) that described the study purpose, time commitment, and location. Recruitment materials followed established guidance for social-media-based recruitment by emphasizing investigator transparency, avoiding targeted outreach that could compromise privacy, and ensuring that participation was clearly voluntary. \cite{Gelinas2017SocialMediaRecruitment}
All sessions were conducted \emph{in person} in an office setting, where the physical sock stimuli were available on-site. Eligibility was restricted to adults (\(\ge 18\) years old). Upon arrival, participants received an information sheet describing the study tasks, data collected, and withdrawal rights, and provided informed consent prior to participation. The study procedures were designed to align with general principles for human research ethics, including respect for autonomy and participant welfare. No monetary payment was provided. Instead, at the conclusion of the session, each participant selected a pair of socks from the available inventory as a thank-you gift. Using a small non-monetary incentive is consistent with research-ethics guidance emphasizing that incentives should not be so valuable as to create undue influence, and that participation can be ethically acceptable without payment when risks are minimal, and consent is robust.

\subsubsection{Estimating mismatch sensitivity via pairwise comparisons}
\label{sec:exp_chi}
For the in-person experiment, each of the \(k\) appearance metrics (e.g., color family, pattern class, length class, material class) took exactly three possible values. Therefore, the number of distinct sock \emph{types} in the full factorial stimulus space is \(|\mathcal{U}| = 3^{k}\). To ensure participants could form both matched and mismatched pairs during the tasks and still receive socks afterward, we stocked at least two physical instances per sock type (i.e., at least \(2\cdot 3^{k}\) individual socks in total), replenishing common choices as needed across sessions.  

Participants complete a paired-comparison task. On each trial \(t\), the participant is shown two candidate \emph{pairs} of socks, \(P_t^A=(s_{t,1}^A,s_{t,2}^A)\) and \(P_t^B=(s_{t,1}^B,s_{t,2}^B)\), and is asked:
\begin{quote}
\say{Which pair would you feel is \emph{less awkward} to wear in public today?}
\end{quote}
Let \(y_{r,t}=1\) if participant \(r\) selects option \(A\) (and \(0\) otherwise). This paradigm follows the classical method of paired comparisons for preference estimation \cite{BradleyTerry1952Paired}.

We model each pair \(P\) by its mismatch severity \(m(P)=g(\eta(P))\), where \(g\) is the same monotone penalty shape used in Eq.~(\ref{eq:social_cost}) (we use \(g(x)=x^\gamma\) in simulation and estimation, with \(\gamma\ge 1\)). The probability that participant \(r\) chooses pair \(A\) over \(B\) is
\begin{equation}
\Pr[y_{r,t}=1]
=
\sigma\!\left(\chi_r \cdot \big(m(P_t^B)-m(P_t^A)\big)\right),
\label{eq:chi_logit}
\end{equation}
where \(\sigma(z)=(1+e^{-z})^{-1}\) is the logistic sigmoid and \(\chi_r\ge 0\) is the participant-specific sensitivity. Intuitively, larger \(\chi_r\) implies more deterministic avoidance of mismatch, whereas smaller \(\chi_r\) implies indifference or noise. We estimate \(\chi_r\) by maximum likelihood over the participant's trials, optionally with a weak regularizer to stabilize estimates for participants with near-random choices. 

\review{In addition to the preference-estimation tasks, we recorded a binary behavioral compliance outcome for each participant. Compliance was defined as whether the participant agreed to wear the assigned visibly non-matching sock pair in the public/office setting at the end of the session. Formally, for participant \(r\), we define \(c_r=1\) if the participant complied with this task and \(c_r=0\) otherwise. The correlations reported in the Results section use this binary variable: \(\mathrm{Corr}(\hat{\chi},c)\), \(\mathrm{Corr}(\hat{\delta},c)\), and \(\mathrm{Corr}(\hat{\chi},\hat{\delta})\).}

\subsubsection{Estimating diversity preference via bundle choices}
\label{sec:exp_delta}
To identify \(\delta\), we construct sock \emph{bundles} \(B\subset\mathbb{S}\) of fixed size \(|B|=M\) and fixed total price (or normalized price), while varying their diversity \(D(B)\). Concretely, we generate bundles spanning a grid of diversity levels by sampling socks from mixture distributions over appearance categories. Low-diversity bundles concentrate on few categories (high cross-compatibility), while high-diversity bundles spread mass across many categories.

Participants are shown a small set of candidate bundles and asked to select the one they would prefer to own for daily use under a fixed time horizon. To reduce hypothetical bias, participants are informed that the study will use their choices to determine the distribution from which their assigned socks are drawn (subject to budget and availability constraints). This design encourages participants to express genuine preference over diversity vs.\ matchability.

We assume that participant \(r\) assigns utility to a bundle \(B\) as
\begin{equation}
U_r(B)=\delta_r \cdot D(B)\;-\;\widehat{C}_{\text{soc}}(B;\widehat{\chi}_r)\;-\;\widehat{C}_{\text{rep}}(B),
\label{eq:delta_utility}
\end{equation}
where \(\delta_r\ge 0\) is the diversity preference to be estimated, \(\widehat{\chi}_r\) is the participant's mismatch sensitivity estimate, \(\widehat{C}_{\text{soc}}(B;\widehat{\chi}_r)\) is an estimated expected social penalty induced by the bundle (computed by simulating daily pairing within \(B\) under a reference exposure rate), and \(\widehat{C}_{\text{rep}}(B)\) is an estimated replacement propensity term (e.g., expected orphaning / stranded capacity under a reference loss regime). Participant choices over bundles are modeled using a random-utility multinomial logit model \cite{McFadden1974}. We estimate \(\delta_r\) by maximum likelihood \cite{chernoff2011use}.

\review{Importantly, each participant completed \(N_{\chi}=7\) pairwise-comparison trials for estimating mismatch sensitivity and \(N_{\delta}=3\) bundle-choice tasks for estimating diversity preference. In each bundle-choice task, participants selected one preferred bundle from \(J=10\) candidate bundles, each containing \(M=6\) socks. The sock stimuli were defined by \(k=4\) appearance dimensions, each with three possible values, yielding \(3^4=81\) possible sock types. Trial order was randomized independently for each participant, and the left/right order of pairwise-comparison alternatives, as well as the display order of candidate bundles, was randomized within each task. The estimates \(\hat{\chi}\) and \(\hat{\delta}\) were obtained by minimizing the corresponding negative log-likelihoods with weak \(L_2\) regularization. Specifically, we used \(\lambda_{\chi}=0.01\) for mismatch-sensitivity estimation and \(\lambda_{\delta}=0.01\) for diversity-preference estimation. Parameter uncertainty was estimated using \(B=3\) non-parametric bootstrap resamples over each participant's trials/tasks. The full sock-stimulus encoding and the generated pairwise-comparison and bundle-choice task files are provided as supplementary material.}

\review{Importantly, in the empirical and simulation components, the abstract objects \(\eta\), \(g\), and \(D(S)\) are instantiated using the visible sock-stimulus encoding. Each sock type is represented by a categorical appearance vector \(\overrightarrow{u}\), whose entries describe directly observable dimensions such as color family, pattern, length, and material/style class. We define appearance dissimilarity using normalized Hamming distance \(\eta(s_i,s_j)=\frac{1}{k}\sum_{\ell=1}^{k}\mathbb{I}\!\left[u_{i,\ell}\neq u_{j,\ell}\right]\) such that \(\xi(s_i,s_j)=1-\eta(s_i,s_j)\). This metric is used as a transparent operational measure of visible mismatch rather than as a complete psychological model of perception. Perceived mismatch cost is calibrated through the participant-specific sensitivity parameter \(\chi_r\), estimated from pairwise-comparison choices. Thus, two participants can face the same visual dissimilarity \(\eta\) but assign different social costs to it. For the baseline analysis, we use \(g(\eta)=\eta^{1.02}\), which is nearly linear and therefore avoids imposing a strong convex or threshold response by assumption. Diversity \(D(S)\) is measured using Shannon diversity over sock types in the participant's bundle or simulated inventory. To verify that the results are not driven by these specific modeling choices, we also evaluate alternative penalty shapes \(g(\eta)=\eta^2\) and \(g(\eta)=\mathbb{I}[\eta>0]\), and an alternative diversity functional based on mean pairwise feature dispersion. These robustness checks are reported in Table~\ref{tab:user_robustness_transposed}.}

\section{Results}
\label{sec:results}
The results are divided into two parts: user experiment and simulation experiment. The user experiments contribute both direct indicators for the proposed non-matching socks pairing practice as well as providing an estimation for the unknown functions in the model, later used by the simulator to investigate a wider \say{what if} scenarios. 

\subsection{User experiment}
\label{sec:results_user}
\review{Overall, the study included \(N=60\) adult participants. The participants' ages ranged from 18 to 62, with a mean \(\pm\) standard deviation of \(37.1 \pm 6.8\) years. Of the participants, 43 were female (\(71.7\%\)) and 17 were male (\(28.3\%\)).} The mean estimated mismatch sensitivity was $\overline{\hat{\chi}}=1.120$ (median $0.968$) and the mean estimated diversity preference was $\overline{\hat{\delta}}=1.753$ (median $1.368$). \review{The empirical correlations were \(\mathrm{Corr}(\hat{\chi},c)=0.038\), \(\mathrm{Corr}(\hat{\delta},c)=0.254\), and \(\mathrm{Corr}(\hat{\chi},\hat{\delta})=0.021\), where \(c\) is the binary compliance outcome.}

\begin{figure}[!ht]
\centering
\begin{subfigure}{0.33\linewidth}
\centering
\includegraphics[width=\linewidth]{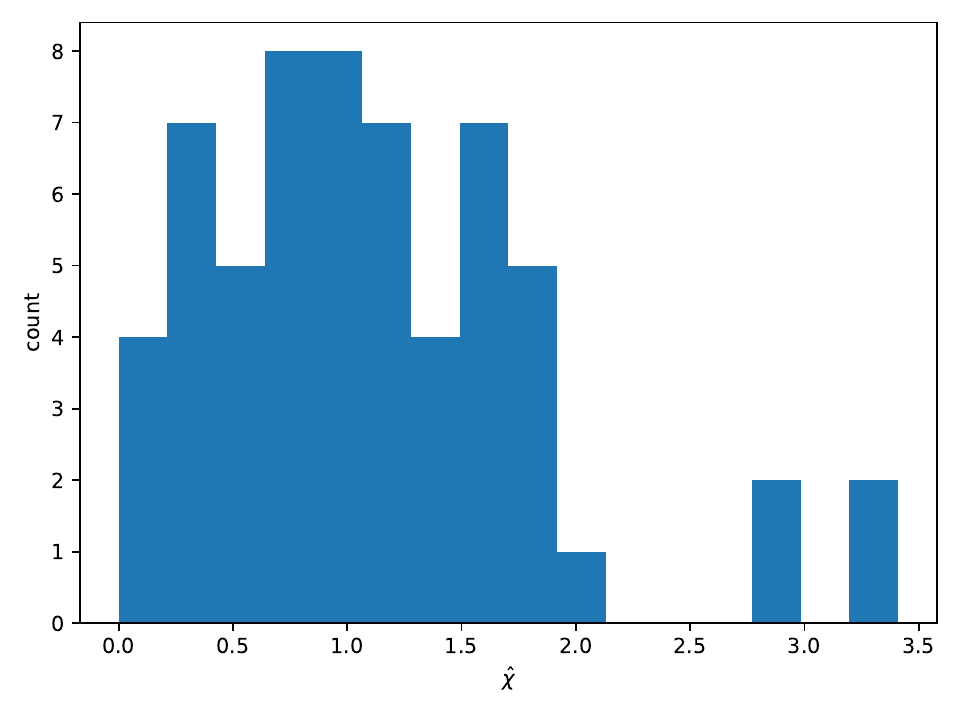}
\caption{Histogram of $\hat{\chi}$.}
\label{fig:user_chi_hist}
\end{subfigure}\hfill
\begin{subfigure}{0.33\linewidth}
\centering
\includegraphics[width=\linewidth]{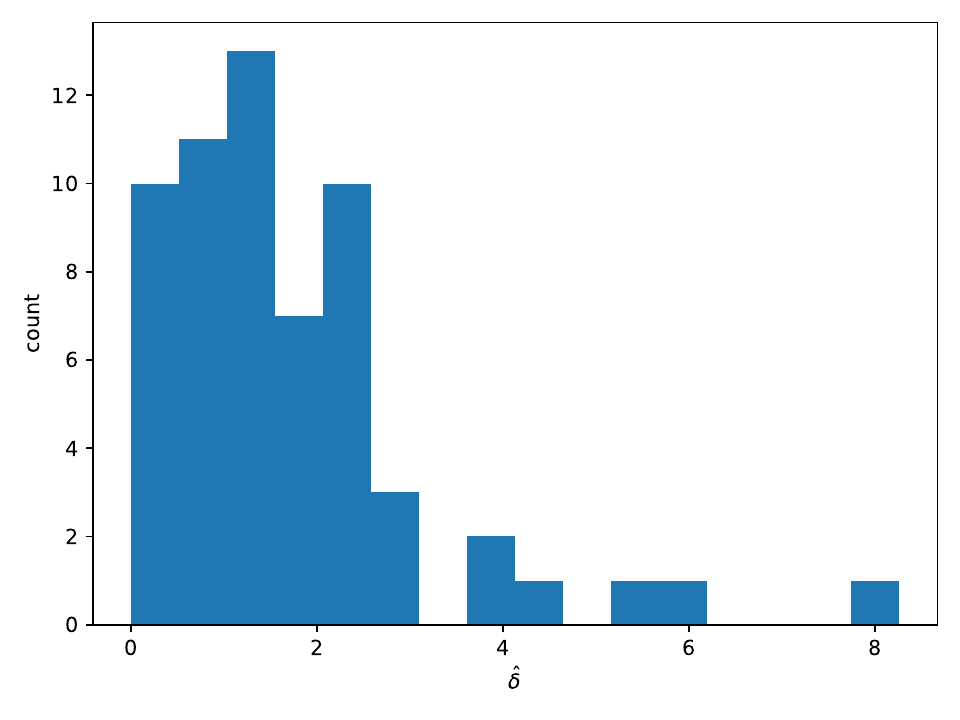}
\caption{Histogram of $\hat{\delta}$.}
\label{fig:user_delta_hist}
\end{subfigure}
\begin{subfigure}{0.33\linewidth}
\centering
\includegraphics[width=\linewidth]{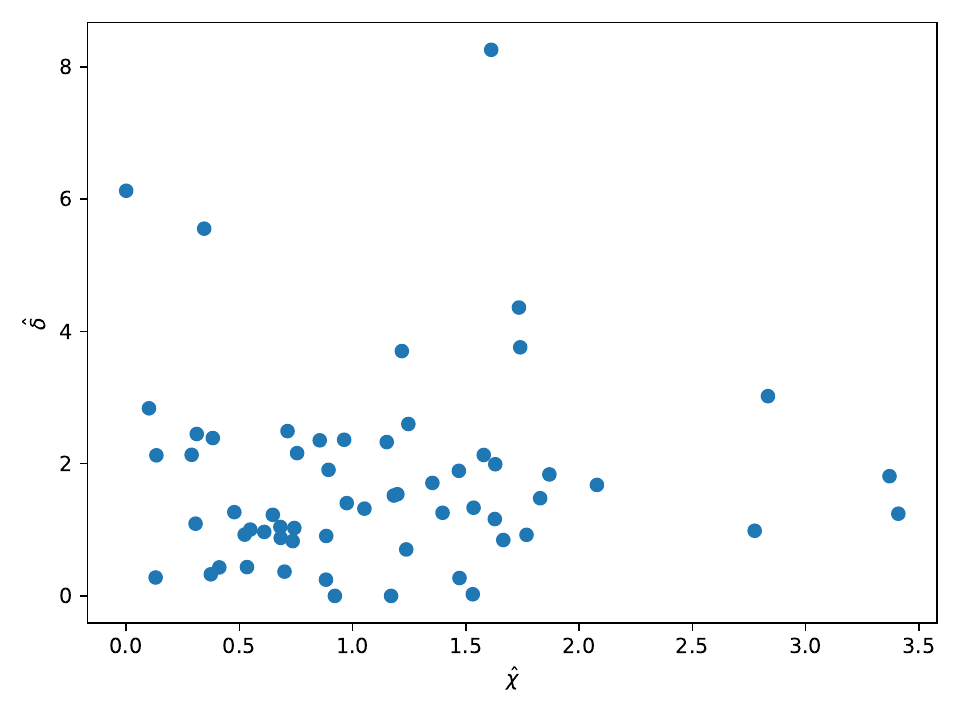}
\caption{$\hat{\chi}$ vs.\ $\hat{\delta}$.}
\label{fig:user_chi_delta}
\end{subfigure}\hfill
\caption{User-study parameter estimates and feasibility outcomes.}
\label{fig:user_study}
\end{figure}

Figure~\ref{fig:user_study} outlines the user-study parameter estimates and feasibility outcomes. In particular, Figure~\ref{fig:user_chi_hist} shows substantial heterogeneity in mismatch sensitivity, with most estimates concentrated in the range $\hat{\chi}\in[0,2]$ and a smaller right tail extending to $\hat{\chi}= 3.4$. Operationally, low $\hat{\chi}$ participants are relatively indifferent to mismatch (or noisy in pairwise comparisons), while high $\hat{\chi}$ participants exhibit strong aversion to visually non-matching pairs, implying a larger expected social penalty under public exposure (Eq.~\ref{eq:social_cost}). Figure~\ref{fig:user_delta_hist} indicates that diversity preference is also right-skewed: most participants cluster around modest values ($\hat{\delta}= 0.5$--$2.5$), while a small number of participants exhibit very high diversity preference (up to $\hat{\delta}= 8.3$). This pattern supports modeling $\delta$ as a heterogeneous trait: for many participants, matchability dominates bundle utility, but a minority substantially values variety even at the expense of increased orphaning risk. Figure~\ref{fig:user_chi_delta} plots $\hat{\delta}$ against $\hat{\chi}$. The relationship is visually diffuse and the empirical correlation is near zero (Corr$(\hat{\chi},\hat{\delta})=0.021$), suggesting that mismatch aversion and diversity-seeking are largely independent in this sample. This motivates treating $\chi$ and $\delta$ as separate behavioral parameters in simulation rather than collapsing them into a single \say{style} axis.

\review{To assess estimation uncertainty and model adequacy, we computed standard errors and 95\% confidence intervals for the participant-level estimates \(\hat{\chi}\) and \(\hat{\delta}\). Model fit was evaluated using the maximized log-likelihood, McFadden's pseudo-\(R^2\), and choice-prediction accuracy relative to a random-choice baseline. We further conducted a validation check by re-estimating the choice models under held-out trials/tasks and measuring whether the fitted model predicted the held-out choices above chance. Finally, we tested robustness to alternative mismatch-penalty shapes \(g(\eta)\in\{\eta,\eta^2,\mathbb{I}[\eta>0]\}\) and alternative diversity measures, including Shannon diversity and mean pairwise feature dispersion. The qualitative conclusions were unchanged across these specifications: \(\hat{\chi}\) and \(\hat{\delta}\) remained heterogeneous, their mutual correlation remained close to zero, and the correlation between \(\hat{\delta}\) and compliance remained larger than the correlation between \(\hat{\chi}\) and compliance.}

\review{Table~\ref{tab:user_robustness_transposed} reports the robustness analysis for the user-study estimates under alternative mismatch-penalty shapes and diversity measures. The estimates are stable across specifications. The mean mismatch-sensitivity estimate remains in the range \(0.742\)--\(1.113\), and the mean diversity-preference estimate remains in the range \(1.619\)--\(2.287\). Across all specifications, \(\mathrm{Corr}(\hat{\chi},\hat{\delta})\) remains close to zero, indicating that mismatch aversion and diversity preference capture distinct behavioral dimensions. In addition, the association between diversity preference and compliance is consistently positive and larger than the corresponding association for mismatch sensitivity. Held-out prediction accuracy for the pairwise-comparison task is above chance in all specifications, while bundle-choice prediction is more modest, consistent with the larger choice set in the bundle task. Overall, the robustness analysis supports the main interpretation that the user-study parameters capture heterogeneous but distinct behavioral traits, and that the substantive conclusions are not driven by a single penalty-shape or diversity-measure choice.}

\begin{table}[!ht]
\centering
\small
\begin{tabular}{lcccc}
\hline \hline
Measure 
& Baseline: $g(\eta)=\eta^{1.02}$
& Convex: $g(\eta)=\eta^2$
& Threshold: $g(\eta)=\mathbb{I}[\eta>0]$
& Baseline penalty, mean pairwise diversity \\
\hline \hline
Mean $\hat{\chi}$ 
& 1.113 & 0.997 & 0.742 & 1.113 \\
Mean $\hat{\delta}$ 
& 1.728 & 1.619 & 1.851 & 2.287 \\
Corr$(\hat{\chi},c)$ 
& 0.038 & 0.061 & -0.059 & 0.038 \\
Corr$(\hat{\delta},c)$ 
& 0.228 & 0.225 & 0.219 & 0.245 \\
Corr$(\hat{\chi},\hat{\delta})$ 
& 0.032 & -0.033 & 0.114 & 0.009 \\
Held-out $A_{\chi}$ 
& 0.591 & 0.585 & 0.545 & 0.591 \\
Held-out $A_{\delta}$ 
& 0.288 & 0.300 & 0.288 & 0.233 \\
\hline \hline
\end{tabular}
\caption{Robustness of the user-study estimates to alternative mismatch-penalty shapes and diversity measures. \(c\) denotes the binary compliance variable. Held-out \(A_{\chi}\) is pairwise-comparison prediction accuracy on held-out trials, and held-out \(A_{\delta}\) is bundle-choice prediction accuracy on held-out tasks.}
\label{tab:user_robustness_transposed}
\end{table}

\subsection{Simulations}

\subsubsection{Exact benchmark on tractable instances}
\review{Table~\ref{tab:exact_benchmark_avg} reports the exact benchmark results for small finite-horizon instances, averaged across the three benchmark settings. The purpose of this analysis is not to solve the full-scale sock-planning problem exactly, but to determine whether the interpretable policies are reasonable approximations when the optimum is known. Greedy, ThresholdMix, and ExposureAware closely match the exact dynamic-programming optimum, with an average objective value of $6.840 \pm 0.868$ compared with $6.840 \pm 0.868$ for the optimum, and an average optimality gap of only $0.009 \pm 0.007\%$. These policies also reproduce the optimal infeasibility level, with $0.483 \pm 0.633$ infeasible days, and the same stranded capacity, $10.967 \pm 2.237$. In contrast, Purist obtains a lower average objective value of $5.589 \pm 0.872$ and a substantially larger optimality gap of $17.563 \pm 14.618\%$, mainly because strict identical matching increases infeasible days to $1.205 \pm 1.248$ and stranded capacity to $12.411 \pm 2.869$. OrphanRescue avoids the additional infeasibility of Purist, but its average social cost is much higher ($0.997 \pm 0.414$), producing an average optimality gap of $11.594 \pm 2.794\%$. Overall, the exact benchmark supports interpreting Greedy, ThresholdMix, and ExposureAware as near-optimal practical heuristics on small instances, while also clarifying that Purist and OrphanRescue fail for different reasons: the former is too restrictive, whereas the latter accepts too much mismatch.}

\begin{table}[!ht]
\centering
\small
\begin{tabular}{lccccc}
\hline \hline
\textbf{Policy} & \textbf{Objective value} & \textbf{Gap (\%)} & \textbf{Infeasible days} & $C_{\text{soc}}$ & $C_{\text{strand}}$ \\
\hline \hline
Optimal       & $6.840 \pm 0.868$ & $0.000 \pm 0.000$  & $0.483 \pm 0.633$ & $0.193 \pm 0.156$ & $10.967 \pm 2.237$ \\
Purist        & $5.589 \pm 0.872$ & $17.563 \pm 14.618$ & $1.205 \pm 1.248$ & $0.000 \pm 0.000$ & $12.411 \pm 2.869$ \\
Greedy        & $6.840 \pm 0.868$ & $0.009 \pm 0.007$  & $0.483 \pm 0.633$ & $0.194 \pm 0.157$ & $10.967 \pm 2.237$ \\
ThresholdMix  & $6.840 \pm 0.868$ & $0.009 \pm 0.007$  & $0.483 \pm 0.633$ & $0.194 \pm 0.157$ & $10.967 \pm 2.237$ \\
OrphanRescue  & $6.036 \pm 0.664$ & $11.594 \pm 2.794$ & $0.483 \pm 0.633$ & $0.997 \pm 0.414$ & $10.967 \pm 2.237$ \\
ExposureAware & $6.840 \pm 0.868$ & $0.009 \pm 0.007$  & $0.483 \pm 0.633$ & $0.194 \pm 0.157$ & $10.967 \pm 2.237$ \\
\hline \hline
\end{tabular}
\caption{Exact optimization benchmark on small tractable instances, averaged across the three benchmark instances. Values are reported as mean $\pm$ standard deviation. The optimal policy is computed by finite-horizon dynamic programming/backward induction; heuristic policies are evaluated on the same instances. The gap is reported relative to the exact optimal objective value.}
\label{tab:exact_benchmark_avg}
\end{table}

\subsubsection{Reference-regime simulation}
Table~\ref{tab:e1_csv} compares policies under the reference initialization\review{, which is larger than the exact benchmark instances and therefore evaluated by Monte Carlo simulation rather than exhaustive dynamic programming.}. The Purist strategy achieves zero social cost by construction, but suffers from a large number of infeasible days (mean of 88) because strict identical matching interacts poorly with stochastic availability (laundry buffering) and gradual attrition. Allowing non-identical pairing dramatically improves feasibility: Greedy, ThresholdMix, and OrphanRescue reduce infeasible days by roughly $66$--$74$ days relative to Purist. Among the mixing-aware policies, ThresholdMix attains low social cost ($C_{\text{soc}} = 2.7$) with strong feasibility, while OrphanRescue attains similar feasibility at a substantially higher social penalty ($C_{\text{soc}} = 47.8$), reflecting that \say{rescuing} low-degree socks sometimes requires visibly mismatched pairings.

\begin{table}[!ht]
\centering
\begin{tabular}{lcccccc}
\hline \hline
Policy & \# socks & $C_{\$}$ & $C_{\text{eco}}$ & $C_{\text{soc}}$ & Infeasible days & $C_{\text{strand}}$ \\
\hline \hline
Purist & 29.48 $\pm$ 0.74 & 200.00 $\pm$ 0.00 & 200.00 $\pm$ 0.00 & 0.00 $\pm$ 0.00 & 87.9 $\pm$ 12.4 & 919.88 $\pm$ 25.79 \\
Greedy & 31.37 $\pm$ 0.74 & 199.07 $\pm$ 0.14 & 199.07 $\pm$ 0.14 & 3.53 $\pm$ 0.68 & 21.8 $\pm$ 6.5 & 882.05 $\pm$ 31.81 \\
ThresholdMix & 33.69 $\pm$ 0.79 & 200.00 $\pm$ 0.00 & 200.00 $\pm$ 0.00 & 2.73 $\pm$ 0.63 & 14.7 $\pm$ 5.1 & 983.87 $\pm$ 35.74 \\
OrphanRescue & 35.31 $\pm$ 0.81 & 200.00 $\pm$ 0.00 & 200.00 $\pm$ 0.00 & 47.84 $\pm$ 2.12 & 13.8 $\pm$ 4.7 & 1063.13 $\pm$ 36.34 \\
\hline\hline
\end{tabular}
\caption{Reference-regime policy comparison (mean $\pm$ 95\% CI half-width).}
\label{tab:e1_csv}
\end{table}

Table~\ref{tab:e2_csv} reports mean outcomes over a grid of disappearance probabilities $d$ and wear limits $\theta$. Two qualitative patterns appear. First, when socks wear out quickly (e.g., $\theta=15$), all policies exhibit a high infeasibility rate, which increases further as $d$ grows; in such regimes, the system is dominated by replacement pressure and stochastic unavailability. Second, when socks last longer (e.g., $\theta=40$) and loss is negligible ($d\approx 0$), infeasible days are near zero for all policies; differences then manifest primarily in secondary outcomes (e.g., stranded capacity), reflecting different inventory and pairing dynamics.

\begin{longtable}{cccccc}
\hline \hline 
$\theta$ & $d$ & Policy & $\mathbb{E}[\#\text{socks}]$ & $\mathbb{E}[\text{infeasible days}]$ & $\mathbb{E}[C_{\text{strand}}]$ \\
\hline \hline 
\endfirsthead

\hline \hline 
$\theta$ & $d$ & Policy & $\mathbb{E}[\#\text{socks}]$ & $\mathbb{E}[\text{infeasible days}]$ & $\mathbb{E}[C_{\text{strand}}]$ \\
\hline \hline 
\endhead

\hline \hline 
\multicolumn{6}{r}{\emph{Continued on next page}} \\
\hline \hline 
\endfoot

\hline \hline 
\endlastfoot

\multirow{24}{*}{15} & \multirow{4}{*}{0.00} & Purist       & 35.78 & 109.4 & 25.55 \\
                    &                      & Greedy       & 38.13 & 80.6  & 3.17  \\
                    &                      & ThresholdMix & 31.80 & 126.5 & 0.00  \\
                    &                      & OrphanRescue & 39.83 & 107.5 & 82.53 \\
\cline{2-6}
 & \multirow{4}{*}{0.03} & Purist       & 33.80 & 187.3 & 151.60 \\
                    &                      & Greedy       & 34.57 & 161.8 & 107.33 \\
                    &                      & ThresholdMix & 35.37 & 155.6 & 254.32 \\
                    &                      & OrphanRescue & 32.92 & 201.5 & 152.48 \\
\cline{2-6}
 & \multirow{4}{*}{0.06} & Purist       & 37.47 & 212.9 & 548.35 \\
                    &                      & Greedy       & 35.23 & 197.7 & 456.93 \\
                    &                      & ThresholdMix & 37.63 & 185.3 & 559.36 \\
                    &                      & OrphanRescue & 33.57 & 229.0 & 496.15 \\
\cline{2-6}
 & \multirow{4}{*}{0.09} & Purist       & 37.32 & 250.1 & 903.95 \\
                    &                      & Greedy       & 33.87 & 238.5 & 796.07 \\
                    &                      & ThresholdMix & 30.85 & 252.8 & 771.68 \\
                    &                      & OrphanRescue & 26.22 & 282.6 & 572.10 \\
\cline{2-6}
 & \multirow{4}{*}{0.12} & Purist       & 43.73 & 251.6 & 1258.65 \\
                    &                      & Greedy       & 27.98 & 288.7 & 805.92  \\
                    &                      & ThresholdMix & 40.50 & 236.3 & 1419.53 \\
                    &                      & OrphanRescue & 33.07 & 274.2 & 1116.82 \\
\cline{2-6}
 & \multirow{4}{*}{0.15} & Purist       & 36.08 & 263.7 & 1051.60 \\
                    &                      & Greedy       & 29.55 & 302.0 & 876.43  \\
                    &                      & ThresholdMix & 37.48 & 268.2 & 1301.62 \\
                    &                      & OrphanRescue & 30.13 & 304.3 & 891.95  \\
\hline

\multirow{24}{*}{25} & \multirow{4}{*}{0.00} & Purist       & 29.90 & 39.7  & 96.88  \\
                    &                      & Greedy       & 40.70 & 1.6   & 290.68 \\
                    &                      & ThresholdMix & 49.98 & 0.4   & 520.32 \\
                    &                      & OrphanRescue & 46.62 & 5.6   & 446.63 \\
\cline{2-6}
 & \multirow{4}{*}{0.03} & Purist       & 29.08 & 139.2 & 304.62 \\
                    &                      & Greedy       & 42.87 & 55.4  & 492.48 \\
                    &                      & ThresholdMix & 40.75 & 114.5 & 512.75 \\
                    &                      & OrphanRescue & 41.23 & 101.9 & 506.70 \\
\cline{2-6}
 & \multirow{4}{*}{0.06} & Purist       & 33.07 & 228.5 & 553.75 \\
                    &                      & Greedy       & 36.13 & 161.1 & 495.60 \\
                    &                      & ThresholdMix & 37.37 & 150.9 & 505.93 \\
                    &                      & OrphanRescue & 43.92 & 129.9 & 627.82 \\
\cline{2-6}
 & \multirow{4}{*}{0.09} & Purist       & 35.43 & 269.1 & 682.07 \\
                    &                      & Greedy       & 31.57 & 222.4 & 574.38 \\
                    &                      & ThresholdMix & 38.80 & 206.8 & 571.86 \\
                    &                      & OrphanRescue & 37.08 & 222.5 & 606.25 \\
\cline{2-6}
 & \multirow{4}{*}{0.12} & Purist       & 35.48 & 280.0 & 723.02 \\
                    &                      & Greedy       & 33.00 & 262.1 & 631.97 \\
                    &                      & ThresholdMix & 33.38 & 283.0 & 526.61 \\
                    &                      & OrphanRescue & 37.73 & 271.3 & 717.42 \\
\cline{2-6}
 & \multirow{4}{*}{0.15} & Purist       & 37.08 & 292.8 & 782.75 \\
                    &                      & Greedy       & 34.13 & 281.3 & 685.88 \\
                    &                      & ThresholdMix & 26.70 & 306.9 & 551.23 \\
                    &                      & OrphanRescue & 36.52 & 276.6 & 736.08 \\
\hline

\multirow{24}{*}{40} & \multirow{4}{*}{0.00} & Purist       & 32.32 & 0.0   & 562.73  \\
                    &                      & Greedy       & 48.95 & 0.0   & 1228.00 \\
                    &                      & ThresholdMix & 46.62 & 0.0   & 1134.67 \\
                    &                      & OrphanRescue & 49.98 & 0.0   & 1269.33 \\
\cline{2-6}
 & \multirow{4}{*}{0.03} & Purist       & 32.93 & 97.3  & 736.02  \\
                    &                      & Greedy       & 42.88 & 43.6  & 1027.12 \\
                    &                      & ThresholdMix & 41.73 & 59.2  & 945.07  \\
                    &                      & OrphanRescue & 40.70 & 91.2  & 1006.47 \\
\cline{2-6}
 & \multirow{4}{*}{0.06} & Purist       & 35.05 & 189.8 & 882.22 \\
                    &                      & Greedy       & 39.95 & 121.2 & 975.15 \\
                    &                      & ThresholdMix & 39.88 & 123.1 & 960.63 \\
                    &                      & OrphanRescue & 38.70 & 130.1 & 937.35 \\
\cline{2-6}
 & \multirow{4}{*}{0.09} & Purist       & 35.55 & 233.6 & 944.33 \\
                    &                      & Greedy       & 38.72 & 192.3 & 953.58 \\
                    &                      & ThresholdMix & 37.57 & 195.3 & 891.82 \\
                    &                      & OrphanRescue & 37.08 & 198.4 & 912.77 \\
\cline{2-6}
 & \multirow{4}{*}{0.12} & Purist       & 36.08 & 254.8 & 962.45 \\
                    &                      & Greedy       & 36.98 & 240.9 & 910.60 \\
                    &                      & ThresholdMix & 38.12 & 234.5 & 900.57 \\
                    &                      & OrphanRescue & 36.90 & 244.1 & 925.83 \\
\cline{2-6}
 & \multirow{4}{*}{0.15} & Purist       & 36.85 & 275.5 & 1006.83 \\
                    &                      & Greedy       & 35.12 & 270.1 & 920.70  \\
                    &                      & ThresholdMix & 33.78 & 276.0 & 824.70  \\
                    &                      & OrphanRescue & 36.25 & 273.7 & 940.42  \\

\caption{Loss-and-wear stress test. Reported values are Monte Carlo means for each $(\theta,d)$ and policy.}
\label{tab:e2_csv}\\
\end{longtable}

\begin{figure*}[!ht]
\centering
\setlength{\tabcolsep}{2pt}
\renewcommand{\arraystretch}{1}

\begin{tabular}{ccc}
\includegraphics[width=0.325\textwidth]{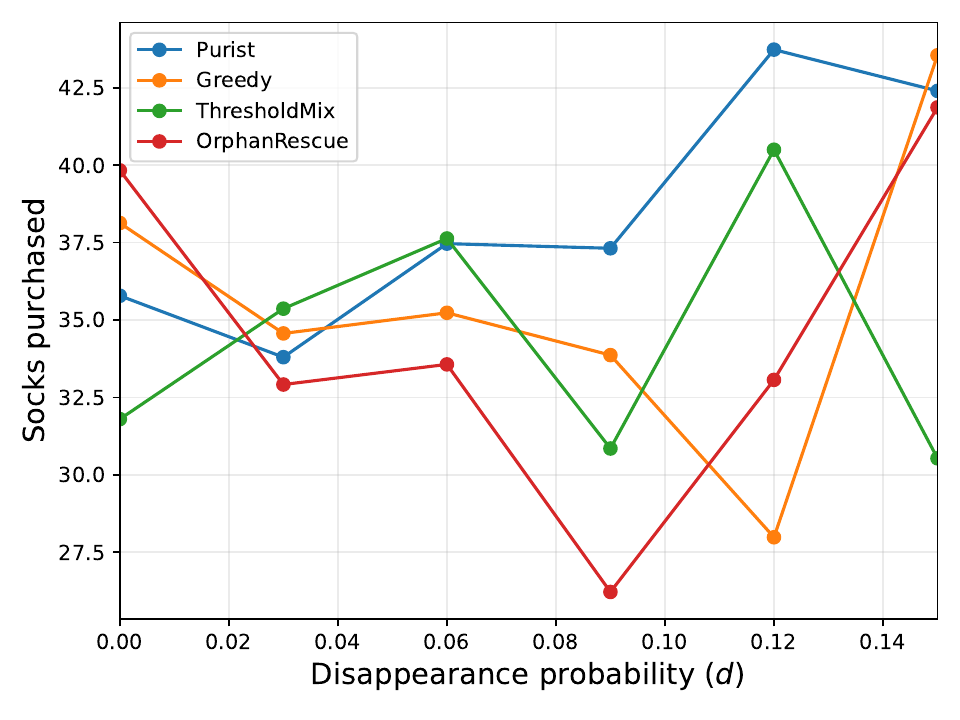} &
\includegraphics[width=0.325\textwidth]{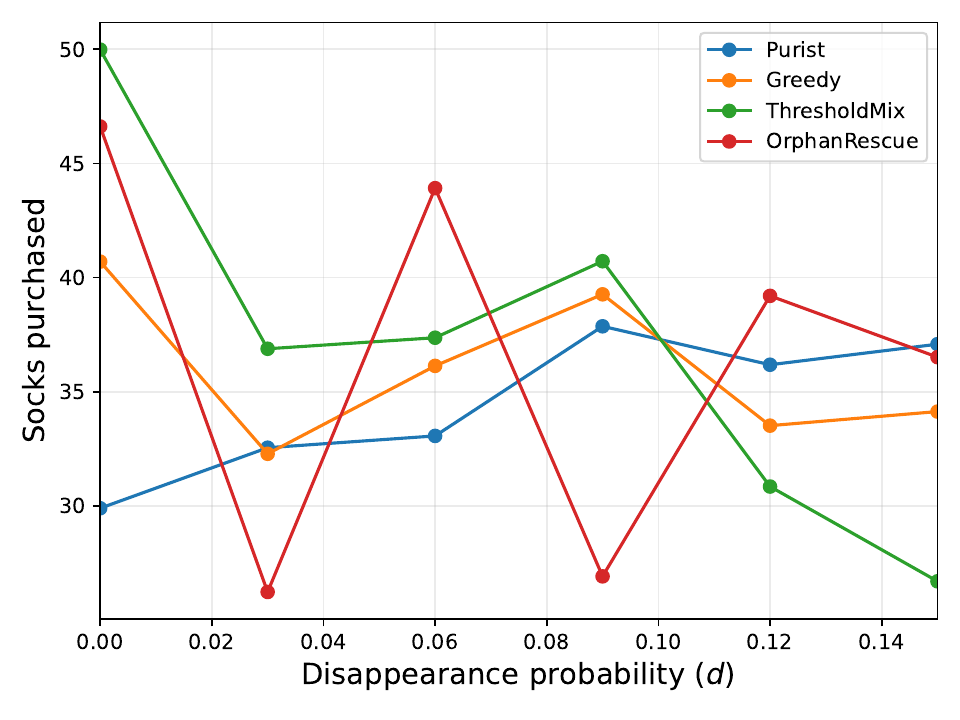} &
\includegraphics[width=0.325\textwidth]{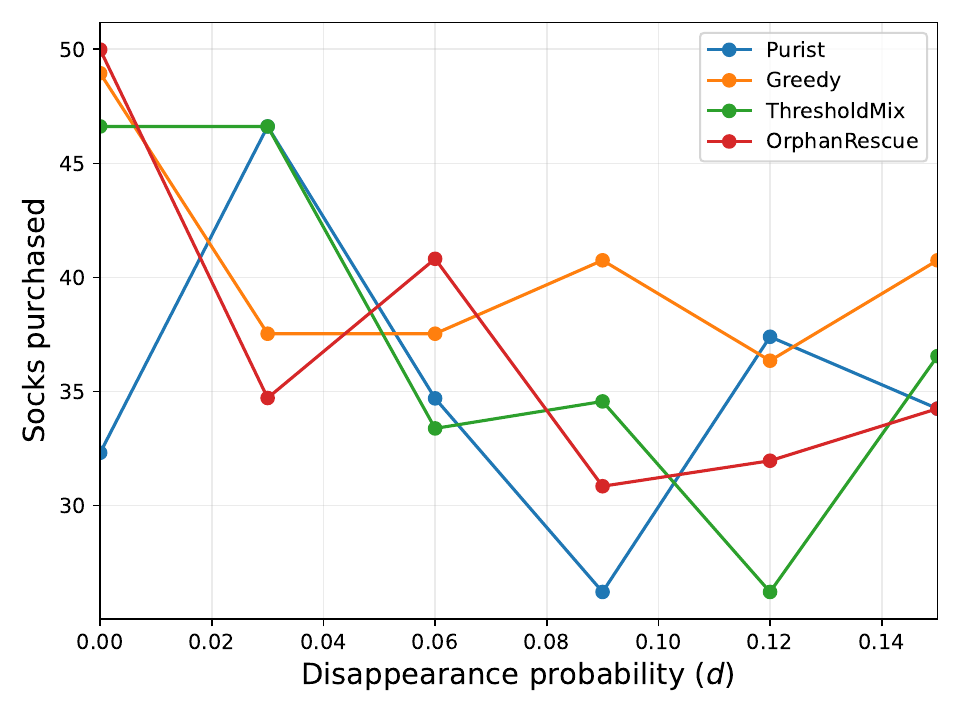} \\

\includegraphics[width=0.325\textwidth]{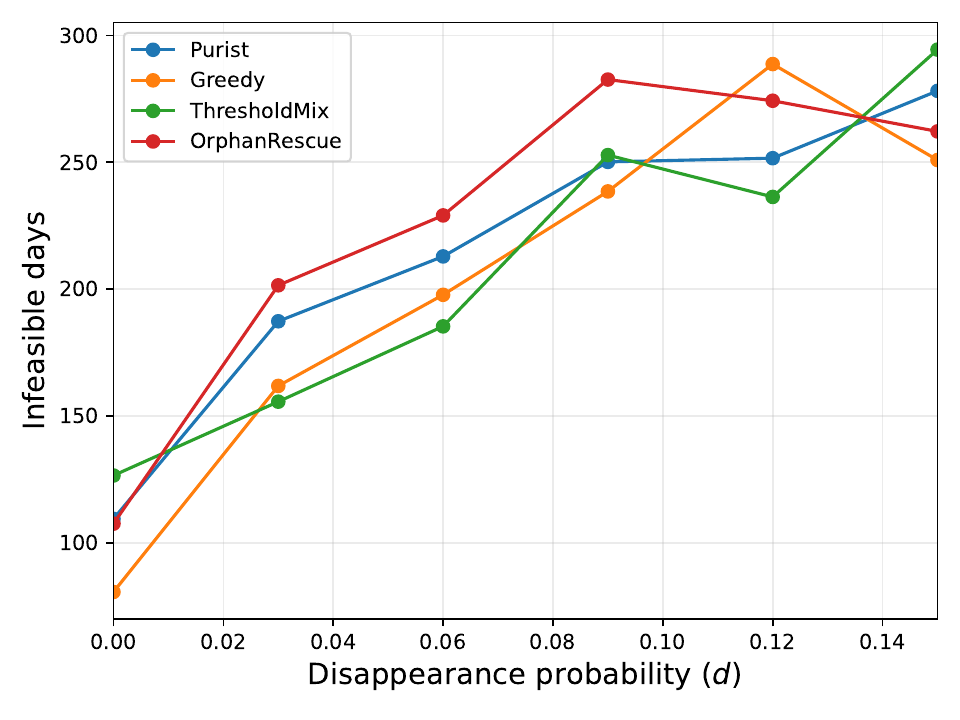} &
\includegraphics[width=0.325\textwidth]{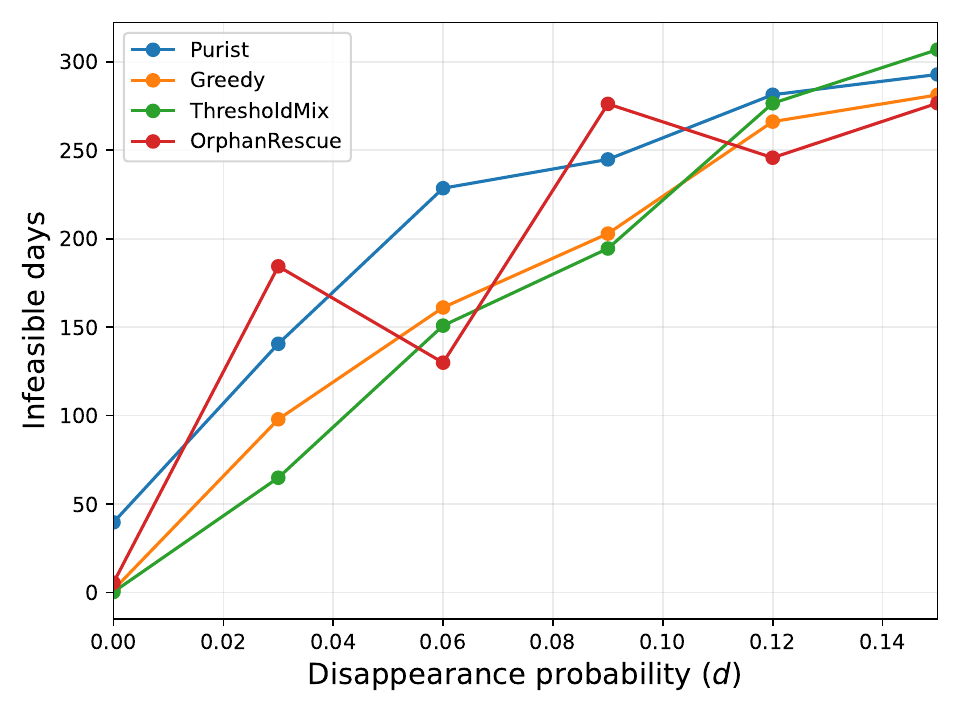} &
\includegraphics[width=0.325\textwidth]{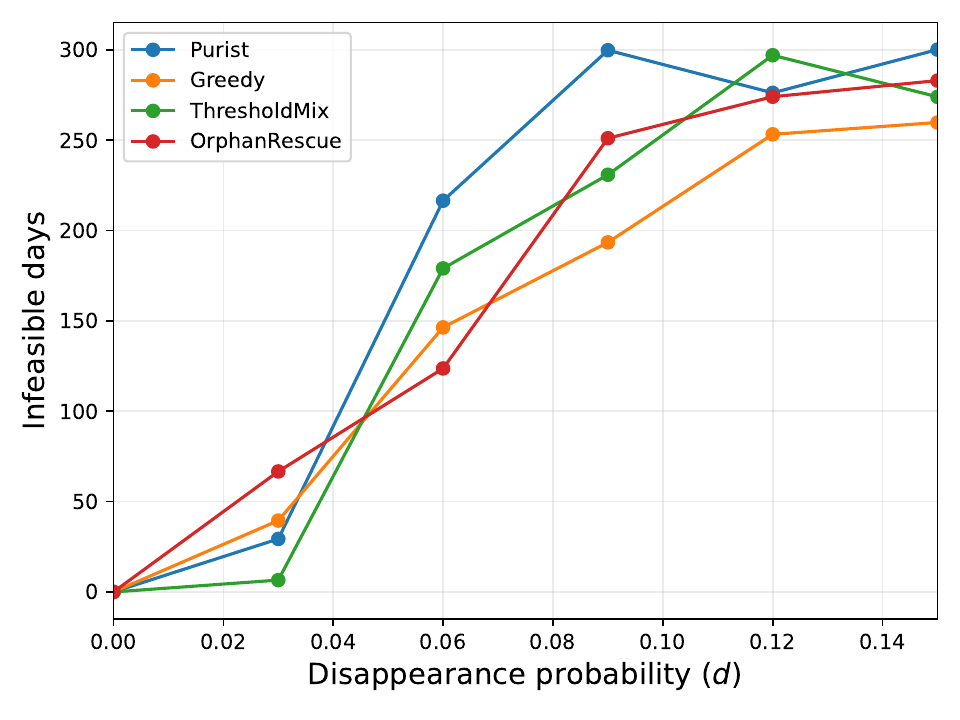} \\

\includegraphics[width=0.325\textwidth]{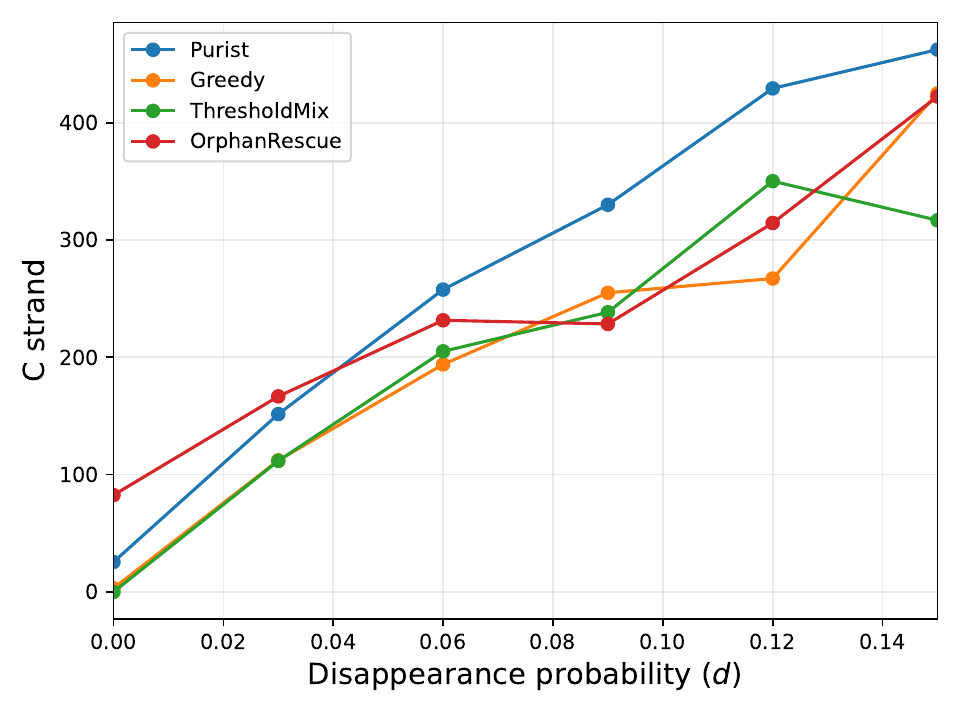} &
\includegraphics[width=0.325\textwidth]{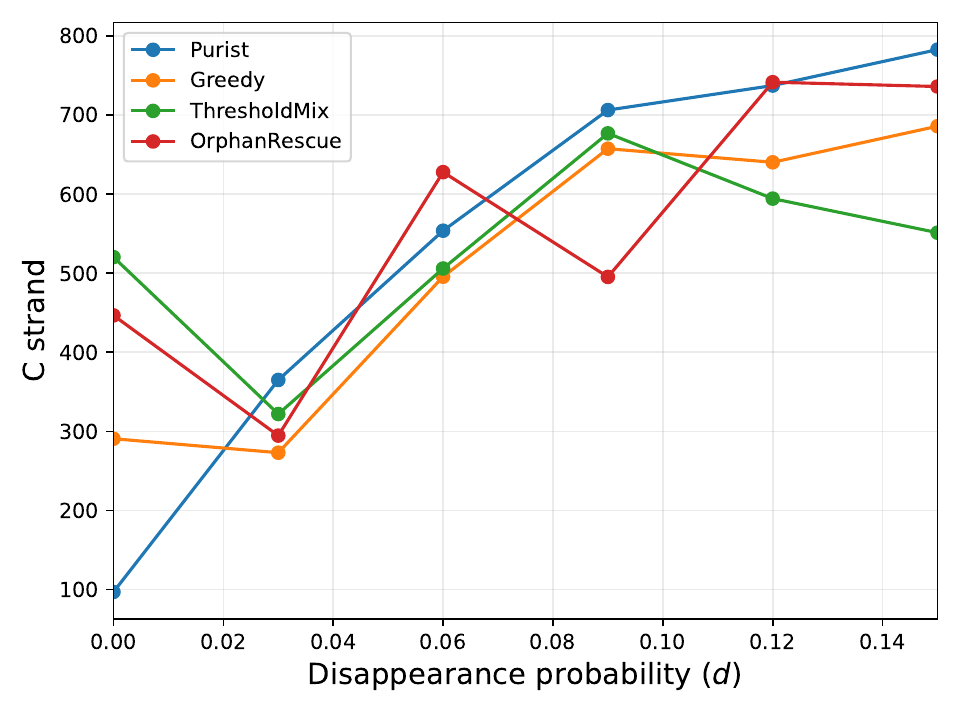} &
\includegraphics[width=0.325\textwidth]{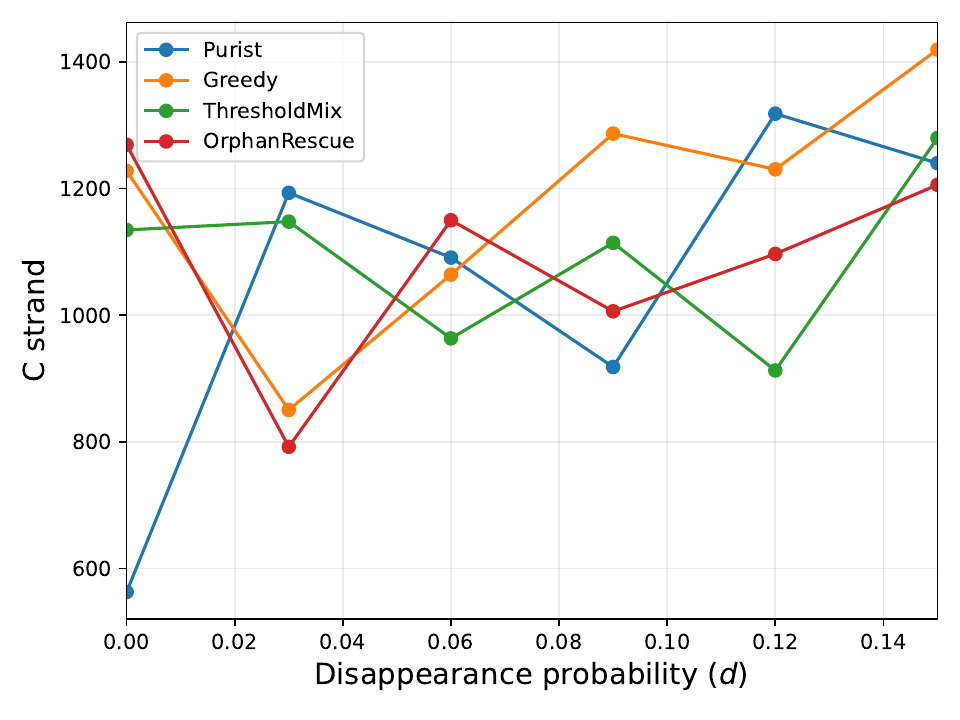} \\
\end{tabular}

\caption{Loss-and-wear stress test. In every panel, the x-axis is the disappearance probability \(d\), and the four curves correspond to the evaluated pairing policies.}
\label{fig:e2_grid}
\end{figure*}

Figure~\ref{fig:e2_grid} visualizes how ecological/economic consumption and \say{service quality} degrade as laundry loss increases. The top row reports the expected number of socks purchased as \(d\) increases. A key interpretive point is that \emph{low purchasing is not automatically good}: a strict matching strategy may appear \say{economical} in some regimes mainly because it becomes frequently infeasible (middle row), i.e., it fails to deliver wearable pairs rather than successfully conserving resources. The middle row, therefore, provides the essential feasibility context - as \(d\) grows, all policies eventually suffer from an increasing number of sockless days, but permissive strategies delay this breakdown by allowing orphan socks to be worn with non-identical partners. Notably, at low-to-moderate loss rates (roughly the left half of the x-axis), mixing-aware policies substantially reduce infeasible days relative to strict matching, demonstrating that the \say{pair constraint} amplifies the practical impact of small stochastic disruptions. Second, the bottom row shows that stranded capacity grows with both \(d\) and \(\theta\): when socks last longer, losing a sock wastes more remaining usable life, making loss events more ecologically costly. This highlights why interventions that increase the utilization of remaining socks (i.e., tolerating mismatch) can yield outsized ecological benefit in long-lifetime regimes, even when the immediate visual mismatch is socially non-zero. 

\newpage

\review{Table~\ref{tab:e3_csv} reports the threshold-sweep analysis for the policies with explicit mismatch-tolerance parameters. Because compatibility is computed using normalized Hamming similarity over discrete appearance features, the effective compatibility thresholds are \(\tau_{\xi}\in\{1.00,0.66,0.33,0.00\}\). Lower values of \(\tau_{\xi}\) allow increasingly dissimilar socks to be paired. The results show the expected trade-off between feasibility and social cost. For ThresholdMix, relaxing the threshold reduces infeasible days from \(51.9\pm77.7\) to \(23.5\pm47.3\) and reduces stranded wear-capacity from \(995.0\pm244.9\) to \(938.3\pm274.6\), while increasing social cost from \(0.00\pm0.00\) to \(2.35\pm3.81\). OrphanRescue displays the same feasibility trend, but incurs substantially higher social cost at intermediate thresholds because it prioritizes socks with few admissible partners, which can require more visible mismatch. Thus, the threshold parameter provides a direct control knob for trading social acceptability against service feasibility and utilization of remaining sock capacity.}

\begin{table}[!ht]
\centering
\small
\begin{tabular}{lccccc}
\hline \hline
Policy & $\tau_{\xi}$ & $C_{\text{soc}}$ & Infeasible days & $C_{\text{strand}}$ & Mismatch rate \\
\hline \hline
\multirow{4}{*}{ThresholdMix} 
 & 1.00 & $0.00 \pm 0.00$ & $51.9 \pm 77.7$ & $995.0 \pm 244.9$ & $0.000 \pm 0.000$ \\
 & 0.66 & $0.24 \pm 0.78$ & $45.1 \pm 72.5$ & $981.5 \pm 253.1$ & $0.004 \pm 0.013$ \\
 & 0.33 & $2.00 \pm 3.39$ & $26.2 \pm 49.8$ & $943.7 \pm 270.5$ & $0.020 \pm 0.033$ \\
 & 0.00 & $2.35 \pm 3.81$ & $23.5 \pm 47.3$ & $938.3 \pm 274.6$ & $0.022 \pm 0.035$ \\
\hline
\multirow{4}{*}{OrphanRescue} 
 & 1.00 & $0.00 \pm 0.00$ & $59.3 \pm 82.0$ & $1009.8 \pm 243.1$ & $0.000 \pm 0.000$ \\
 & 0.66 & $3.91 \pm 3.67$ & $49.0 \pm 71.8$ & $989.3 \pm 253.9$ & $0.063 \pm 0.056$ \\
 & 0.33 & $25.27 \pm 14.50$ & $32.6 \pm 55.3$ & $956.4 \pm 263.7$ & $0.222 \pm 0.100$ \\
 & 0.00 & $11.76 \pm 8.41$ & $24.5 \pm 44.9$ & $940.2 \pm 281.0$ & $0.107 \pm 0.071$ \\
\hline \hline
\end{tabular}
\caption{Threshold-sweep analysis for mismatch-tolerant pairing policies. Values are reported as mean $\pm$ standard deviation over \(n=100\) Monte Carlo replications. Lower \(\tau_{\xi}\) values permit more mismatched pairings. \(C_{\text{soc}}\) denotes accumulated social cost, \(C_{\text{strand}}\) denotes stranded wear-capacity, and mismatch rate denotes the fraction of feasible wearing days on which the selected pair was non-identical.}
\label{tab:e3_csv}
\end{table}

\newpage

\section{Discussion}
\label{sec:discussion}
This study examined a surprisingly consequential everyday choice: whether to discard orphaned socks or to pair them with non-identical partners. We formalized sock acquisition and daily pairing as a planning problem under uncertainty, where socks gradually wear out and may disappear during laundering, and where mismatching imposes a social cost. The model enabled us to quantify the core trade-off of economic-ecological benefit versus social penalty and to evaluate practical policies through Monte Carlo simulation alongside a complementary in-person experiment designed to estimate key behavioral parameters. \review{Importantly, the broader discussion of sock markets and textile consumption should be understood only as motivating context. The evidence-bearing components of the paper are the formal model, the exact benchmark on small instances, the simulation experiments, and the user study.}

The user study provides a direct feasibility signal for the central behavioral assumption of the paper: that at least some people are willing to wear deliberately non-matching socks in public. \review{This suggests that, within the sampled context, \say{orphan pairing} is not merely a hypothetical recommendation but a behavior that some participants were willing to adopt, at least when framed as intentional.} Beyond feasibility, the estimated preference parameters quantify how strongly participants differ in their tolerance. Mismatch sensitivity estimates $\hat{\chi}$ exhibit substantial heterogeneity (Figure~\ref{fig:user_chi_hist}): most participants fall in a low-to-moderate range, while a smaller tail shows strong aversion. This implies that the social cost of mismatch is not uniform: for some users, a mild mismatch is nearly negligible, while for others it is a meaningful deterrent. Importantly, compliance is almost uncorrelated with mismatch sensitivity (Corr$(\hat{\chi},\text{compliance})=0.038$), suggesting that willingness to enact the behavior is not determined solely by social-cost aversion. In contrast, compliance correlates more with diversity preference (Corr$(\hat{\delta},\text{compliance})=0.254$), indicating that participants who value variety (and therefore may already treat socks as expressive rather than purely utilitarian) are more inclined to accept visible nonconformity. Finally, the near-zero correlation between $\hat{\chi}$ and $\hat{\delta}$ (Corr$(\hat{\chi},\hat{\delta})=0.021$; Figure~\ref{fig:user_chi_delta}) supports modeling them as independent behavioral traits: some people both dislike mismatch and enjoy variety, while others neither care about mismatch nor seek diversity. \review{As such, the simulation results should be interpreted as \say{population-dependent}: the same pairing policy can be near-optimal for one user under a given social-cost profile and unacceptable for another.} Concretely, high-$\chi$ individuals are well served by conservative rules (near-purist or high-threshold ThresholdMix), whereas low-$\chi$ and/or high-$\delta$ individuals can accept more aggressive mixing and thus capture larger ecological/service benefits. This is why the paper evaluates interpretable policies with explicit control knobs (e.g., $\tau_{\xi}$). Thus, such knobs can represent either personalized behavior or context-dependent choices.

A key pattern across the simulation stress test (Table~\ref{tab:e2_csv} and Figure~\ref{fig:e2_grid}) is that low purchasing is not automatically good. The Purist strategy sometimes appears economical in socks purchased, but this often coincides with many infeasible (sockless) days, as loss increases. This reveals a mechanism central to the problem: the pair constraint amplifies small stochastic disruptions. Even modest disappearance probabilities can produce periods where usable socks exist but cannot be combined into an acceptable pair, causing the policy to reduce consumption by simply not providing socks on many days. When considered as a whole, the user study suggests that deliberate mismatch is feasible for a majority, but tolerance varies widely, and in a complementary manner, the simulations show that even mild mismatch tolerance can substantially improve service quality under realistic loss, and can reduce wasted wear-capacity, especially for long-lived socks.

\review{Therefore, this study suggests a potential low-effort mechanism for reducing replacement pressure and improving the utilization of already-owned socks. On the individual level, the results can be distilled into simple behavioral heuristics, which translate a complex stochastic planning problem into a rule-of-thumb that can be adopted without computation. On the market level, the study motivates product and retail interventions such as selling intentionally unpaired socks, \say{orphan-compatible} assortments with high cross-matchability, or subscription services that replenish single socks rather than pairs. These implications should be interpreted as mechanism-oriented rather than as a direct environmental accounting: the model shows when mismatch tolerance can reduce premature replacement and stranded capacity, while the actual environmental magnitude depends on product composition, laundering behavior, transport, and disposal pathways.} \review{Thus, the paper does not rely on macroeconomic claims about the sock industry to establish its findings. Those claims motivate scale, whereas the substantive evidence comes from the model, simulations, and behavioral measurements.}

This study is not without limitations. \review{First, the ecological component is proxy-based rather than a full life-cycle assessment. The simulator captures replacement pressure and stranded wear-capacity through \(C_{\text{eco}}\), but does not model fiber composition, manufacturing, transport, laundering behavior, repair, recycling, landfill disposal, or other end-of-life pathways. Thus, the environmental interpretation should be read as suggestive rather than as a direct estimate of emissions, water use, or waste reduction.} Second, the simulator abstracts away multiple real-world factors such as multiple sock drawers, seasonal rotation, fashion contexts, intentional mismatching as a style choice, differential disappearance by material/size, and strategic laundry behavior (e.g., using washing bags). While these simplifications enable tractable simulation and clear mechanisms, they may miss second-order effects. \review{Third, although we include exact optimization benchmarks on small tractable instances, the full-scale model is not solved exactly. The exact benchmark should therefore be interpreted as a validation device for the proposed heuristics, not as evidence that the heuristics are globally optimal in realistic catalogue-scale settings. Fourth, the social-cost component should be interpreted as anticipated awkwardness rather than an observed social penalty. The study does not measure actual workplace evaluations, third-party judgments, or repeated public wearing behavior, so the social implications should be treated as suggestive and context-dependent.} Lastly, participants were recruited and tested in person in Israel. Social norms around appearance, workplace dress codes, and perceived awkwardness of mismatched socks can vary substantially across cultures and subcultures. Consequently, estimates of mismatch sensitivity (\(\chi\)) and exposure (\(\rho\)) derived from this sample may not generalize globally without recalibration.

\review{Taken jointly, the theoretical analysis, simulation framework, and behavioral measurement pipeline suggest that the \say{lost sock} problem is not merely domestic folklore but a structured source of avoidable under-utilization. Even modest willingness to pair non-identical socks, especially when guided by simple threshold-like rules, can extend the useful service obtained from existing sock inventories and reduce replacement pressure under the modeled assumptions. Whether these utilization gains translate into measurable environmental savings depends on product composition, laundering behavior, and disposal pathways, and should be evaluated in future work using full life-cycle assessment data.}





\bibliography{biblio}

@article{intro_1,
title = {Are running socks beneficial for comfort? The role of the sock and sock fiber type on shoe microclimate and subjective evaluations},
journal = {Textile Research Journal},
volume = {91},
number = {15-16},
year = {2021},
pages = {1698-1712},
author = {West, A. M. and Havenith, G. and Hodder, S.},
}

@article{intro_2,
title = {Effects of sock type on foot skin temperature and thermal demand during exercise},
journal = {Ergonomics},
volume = {47},
year = {2004},
pages = {1657-1668},
author = {Purvis, A. and Tunstall, H.},
}

@InProceedings{intro_3,
author="Tomljenovi{\'{c}}, A.
and Skenderi, Z.
and Kraljevi{\'{c}}, I.
and {\v{Z}}ivi{\v{c}}njak, J.",
editor="Msahli, S.
and Debbabi, F.",
title="Durability and Comfort Assessment of Casual Male Socks",
booktitle="Advances in Applied Research on Textile and Materials - IX",
year="2022",
publisher="Springer International Publishing",
pages="210--215",
}

@article{intro_4,
title = {The effect of socks on vertical and anteroposterior ground reaction forces in walking and running},
journal = {The Foot},
volume = {21},
number = {1},
pages = {1-5},
year = {2011},
author = {Blackmore, T. and Ball, N. and Scurr, J.}
}

@article{intro_5,
title = {Continuous Temperature-Monitoring Socks for Home Use in Patients With Diabetes: Observational Study},
journal = {J Med Internet Res},
volume = {20},
number = {12},
pages = {e12460},
year = {2018},
author = {Reyzelman, A. and Koelewyn, K. and Murphy, M. and Shen, X. and Yu, E. and Pillai, R. and Fu, J. and Scholten, H. and Ma, R.}
}

@article{intro_6,
title = {The Sweet Science of Socks},
journal = {New England Review},
volume = {43},
number = {2},
pages = {176-190},
year = {2022},
author = {Miller, B.}
}

@article{socks_dis_2,
author = {Keyes, J.},
year = {2002},
title = {Where It Is That Things Go},
volume = {43},
number = {4},
journal = {The Massachusetts Review},
pages = {676–687}
}

@article{intro_7,
author = {Claudio, L.},
title = {Waste Couture: Environmental Impact of the Clothing Industry},
journal = {Environmental Health Perspectives},
volume = {115},
number = {9},
pages = {A449-A454},
year = {2007},
}

@article{intro_8,
author = {Shi, Y.},
title = {Chinese Immigrant Women Workers' Mediated Negotiations With Constraints On Their Cultural Identities},
journal = {Feminist Media Studies},
volume = {8},
number = {2},
pages = {143-161},
year = {2008},
}

@article{intro_9,
    author = {Yano, C. R.},
    title = "{Wink on Pink: Interpreting Japanese Cute as It Grabs the Global Headlines}",
    journal = {Journal of Asian Studies},
    volume = {68},
    number = {3},
    pages = {681-688},
    year = {2009}
}

@article{intro_10,
author = {Lenz, E.},
title = {Dressing up: Clothing as A Visible Expression of Identity},
journal = {Art Education},
volume = {55},
number = {5},
pages = {25-32},
year = {2002},
}

@article{ouu_review,
title = {Optimization under uncertainty: state-of-the-art and opportunities},
journal = {Computers \& Chemical Engineering},
volume = {28},
number = {6},
pages = {971-983},
year = {2004},
author = {Sahinidis, N. V.}
}

@article{ouu_review_2,
title = {Modeling, analysis, and optimization under uncertainties: a review},
journal = {Structural and Multidisciplinary Optimization},
volume = {64},
pages = {2909–2945},
year = {2021},
author = {Acar, E. and Bayrak, G. and Jung, Y. and Lee, I. and Ramu, P. and Ravichandran, S. S.}
}

@article{ouu_g_1,
title = {A survey on bilevel optimization under uncertainty},
journal = {European Journal of Operational Research},
volume = {311},
pages = {401-426},
year = {2023},
author = {Beck, Y. and Ljubic, I. and Schmidt, M.}
}

@article{ouu_g_2,
title = {Power systems optimization under uncertainty: A review of methods and applications},
journal = {Electric Power Systems Research},
volume = {214},
pages = {108725},
year = {2023},
author = {Roald, L. A. and Pozo, D. and Papavasiliou, A. and Molzahn, D. K. and Kazempour, J. and Conejo, A.}
}

@article{ouu_g_3,
title = {Recent advances in mathematical programming techniques for the optimization of process systems under uncertainty},
journal = {Computers \& Chemical Engineering},
volume = {91},
pages = {3-14},
year = {2016},
author = {Grosmann, I. E. and Apap, R. M. and Calfa, B. A. and Garcia-Herreros, P. and Zhang, Q.}
}

@article{ouu_supply_chain_example,
title = {Energy supply planning and supply chain optimization under uncertainty},
journal = {Journal of Process Control},
volume = {24},
pages = {323-331},
year = {2014},
author = {Lee, J. H.}
}

@article{ouu_hospital,
title = {Data-driven hospitals staff and resources allocation using agent-based simulation and deep reinforcement learning},
journal = {Engineering Applications of Artificial Intelligence},
volume = {126},
pages = {106783},
year = {2023},
author = {Lazebnik, T.}
}

@article{ouu_disaster,
title = {Disaster relief routing under uncertainty: A robust optimization approach},
journal = {IISE Transactions },
volume = {51},
pages = {869-886},
year = {2019},
author = {Li, Y. and Chung, S. H.}
}

@article{ouu_define,
title = {On the treatment of uncertainties in structural mechanics and analysis},
journal = {Computers \& Strucures},
volume = {85},
pages = {235-243},
year = {2007},
author = {Schueller, G. I.}
}

@article{ouu_methods,
  title={Modeling, analysis, and optimization under uncertainties: a review},
  author={Acar, Erdem and Bayrak, Gamze and Jung, Yongsu and Lee, Ikjin and Ramu, Palaniappan and Ravichandran, Suja Shree},
  journal={Structural and Multidisciplinary Optimization},
  volume={64},
  number={5},
  pages={2909--2945},
  year={2021},
  publisher={Springer}
}

@article{ouu_mc,
title = {Dealing with uncertainty: a survey of theories and practices},
journal = {IEEE Trans Knowl Data Eng},
volume = {25},
number = {11},
pages = {2463-2482},
year = {2012},
author = {Li, Y. and Chen, J. and Feng, L.}
}

@article{ouu_exp_1,
title = {Improving accuracy of failure probability estimates with separable Monte Carlo},
journal = {Int J Reliab Saf},
volume = {4},
number = {4},
pages = {393–414},
year = {2010},
author = {Smarslok, B. and Haftka, R. and Carraro, L. and Ginsbourger, D.}
}

@article{ouu_exp_2,
title = {Multilevel Monte Carlo path simulation},
journal = {Oper Res},
volume = {56},
number = {3},
pages = {607–617},
year = {2008},
author = {Giles, M.}
}

@book{bush1994folk,
  author    = {Bush, N.},
  title     = {Folk Socks: The History \& Techniques of Handknitted Footwear},
  year      = {1994},
  publisher = {Interweave Press},
  address   = {Loveland, Colorado}
}

@article{Niinimaki2020FastFashion,
  title        = {The environmental price of fast fashion},
  author       = {Niinim{\"a}ki, Kirsi and Peters, Greg and Dahlbo, Helena and Perry, Patsy and Rissanen, Timo and Gwilt, Alison},
  journal      = {Nature Reviews Earth \& Environment},
  year         = {2020},
  volume       = {1},
  number       = {4},
  pages        = {189--200},
  doi          = {10.1038/s43017-020-0039-9},
  publisher    = {Springer Nature}
}

@article{nguyen2025global,
  title={Global knowledge networks on circular economy and sustainable consumption: Sociological insights into emerging environmental discourses},
  author={Nguyen, Khanh Huy and Tran, Mai Dong and Duong, Trang Thi-Thuy},
  journal={Circular Economy and Sustainability},
  pages={1--30},
  year={2025},
  publisher={Springer}
}

@techreport{Sajn2022TextilesEnvironment,
  title        = {Textiles and the environment},
  author       = {{\v{S}}ajn, Nikolina},
  institution  = {European Parliamentary Research Service (EPRS), European Parliament},
  year         = {2022},
  number       = {PE 729.405},
  month        = {May},
  url          = {https://www.europarl.europa.eu/RegData/etudes/BRIE/2022/729405/EPRS_BRI%282022%29729405_EN.pdf},
  urldate      = {2026-02-20}
}

@book{Goffman1959Presentation,
  title        = {The Presentation of Self in Everyday Life},
  author       = {Goffman, Erving},
  year         = {1959},
  publisher    = {Doubleday},
  address      = {Garden City, NY}
}

@article{AdamGalinsky2012Enclothed,
  title        = {Enclothed cognition},
  author       = {Adam, Hajo and Galinsky, Adam D.},
  journal      = {Journal of Experimental Social Psychology},
  year         = {2012},
  volume       = {48},
  number       = {4},
  pages        = {918--925},
  doi          = {10.1016/j.jesp.2012.02.008}
}

@article{BradleyTerry1952Paired,
  title        = {Rank Analysis of Incomplete Block Designs: I. The Method of Paired Comparisons},
  author       = {Bradley, Ralph Allan and Terry, Milton E.},
  journal      = {Biometrika},
  year         = {1952},
  volume       = {39},
  number       = {3--4},
  pages        = {324--345},
  doi          = {10.1093/biomet/39.3-4.324}
}

@book{GareyJohnson1979,
  title     = {Computers and Intractability: A Guide to the Theory of NP-Completeness},
  author    = {Garey, Michael R. and Johnson, David S.},
  year      = {1979},
  publisher = {W. H. Freeman},
  address   = {San Francisco}
}

@incollection{chernoff2011use,
  title={The use of maximum likelihood estimates in tests for goodness of fit},
  author={Chernoff, Herman and Lehmann, Erich L},
  booktitle={Selected works of EL Lehmann},
  pages={541--549},
  year={2011},
  publisher={Springer}
}

@article{bellezza2014red,
title={The Red Sneakers Effect: Inferring Status and Competence from Signals of Nonconformity},
author={Bellezza, Silvia and Gino, Francesca and Keinan, Anat},
journal={Journal of Consumer Research},
volume={41},
number={1},
pages={35--54},
year={2014},
publisher={Oxford University Press},
doi={10.1086/674870}
}

@article{fortune2025socks,
title={Socks Market Size, Share \& Industry Analysis, 2026-2034},
author={{Fortune Business Insights}},
journal={Market Research Report FBI103875},
year={2025}
}

@article{grandview2025socks,
title={Socks Market Size, Share \& Trends Analysis Report By Product, By Distribution Channel, By Region, And Segment Forecasts, 2026 - 2033},
author={{Grand View Research}},
journal={Industry Analysis Report},
year={2025}
}

@article{gurung2018dressing,
title={Dressing "in code": Clothing rules, propriety, and perceptions},
author={Gurung, Regan AR and Brickner, Michaella and Leet, Mary and Punke, Elizabeth},
journal={The Journal of Social Psychology},
volume={158},
number={5},
pages={553--557},
year={2018},
publisher={Taylor \& Francis},
doi={10.1080/00224545.2017.1393383}
}

@article{slepian2015enclothed,
title={The Cognitive Consequences of Formal Clothing},
author={Slepian, Michael L and Ferber, Simon N and Gold, Joshua M and Rutchick, Abraham M},
journal={Social Psychological and Personality Science},
volume={6},
number={6},
pages={661--668},
year={2015},
publisher={SAGE Publications}
}

@article{KhullerMossNaor1999,
  title   = {The Budgeted Maximum Coverage Problem},
  author  = {Khuller, Samir and Moss, Anna and Naor, Joseph (Seffi)},
  journal = {Information Processing Letters},
  volume  = {70},
  number  = {1},
  pages   = {39--45},
  year    = {1999},
  doi     = {10.1016/S0020-0190(99)00031-9}
}

@misc{gmi2025socks,
  title={Socks Market Analysis},
  author={{Global Market Insights}},
  year={2025},
  url={https://www.gminsights.com/industry-analysis/socks-market}
}

@misc{fmi2025socks,
  title={Socks Market Forecast and Outlook 2026 to 2036},
  author={{Future Market Insights}},
  year={2026},
  url={https://www.futuremarketinsights.com/reports/socks-market}
}

@misc{pmr2025socks,
  title={Socks Market Size, Share, and Growth Forecast, 2025 - 2032},
  author={{Persistence Market Research}},
  year={2025},
  url={https://www.persistencemarketresearch.com/market-research/socks-market.asp}
}

@book{goffman1963stigma,
  title={Stigma: Notes on the Management of Spoiled Identity},
  author={Goffman, Erving},
  year={1963},
  publisher={Prentice-Hall},
  address={Englewood Cliffs, N.J.}
}

@inproceedings{harrison2010introduction,
  title={Introduction to monte carlo simulation},
  author={Harrison, Robert L},
  booktitle={AIP conference proceedings},
  volume={1204},
  pages={17},
  year={2010}
}

@article{lsu2025stigma,
  title={What factors influence SNAP participation? Literature reflecting enrollment in food assistance programs from a social and behavioral science perspective},
  author={Pinard, Courtney A and Bertmann, Farryl MW and Byker Shanks, C and Schober, Daniel J and Smith, Teresa M and Carpenter, Leah C and Yaroch, Amy L},
  journal={Journal of Hunger \& Environmental Nutrition},
  volume={12},
  number={2},
  pages={151--168},
  year={2017},
  publisher={Taylor \& Francis}
}

@book{jinnah2025fashion,
  title={The psychology of fashion},
  author={Mair, Carolyn},
  year={2024},
  publisher={Routledge}
}

@inproceedings{McFadden1974,
  title={Multinomial logit models},
  author={So, Ying and Kuhfeld, Warren F},
  booktitle={SUGI 20 conference proceedings},
  volume={1995},
  pages={1227--1234},
  year={1995}
}

@article{Gelinas2017SocialMediaRecruitment,
  title   = {Using Social Media as a Research Recruitment Tool: Ethical Issues and Recommendations},
  author  = {Gelinas, Luke and Pierce, Robin and Winkler, Sabune and Cohen, I. Glenn and Fernandez Lynch, Holly and Bierer, Barbara E.},
  journal = {The American Journal of Bioethics},
  year    = {2017},
  volume  = {17},
  number  = {3},
  pages   = {3--14},
  doi     = {10.1080/15265161.2016.1276644}
}
\bibliographystyle{unsrt}

\section*{Appendix}

\subsection*{Theoretical analysis}
\label{sec:theory}
The proposed sock matching problem is hard. It is NP-hard and even hard to approximate. In this section, we provide mathematical proofs for these two claims. 

\subsubsection*{This is hard: proving NP-hardness}
We prove that the proposed sock buying-and-matching problem is NP-hard by showing that even a heavily restricted special case is NP-complete. Formally, define the following decision problem.
\begin{definition}[\textsc{Sock-Plan} decision problem]
\label{def:sockplan}
Given: a sock catalogue \(\mathbb{S}\); a price function \(p:\mathbb{S}\to\mathbb{N}\); a dissimilarity (mismatch) function \(\eta:\mathbb{S}\times\mathbb{S}\to[0,1]\) and corresponding compatibility \(\xi(s_i,s_j)=1-\eta(s_i,s_j)\); a horizon \(T\in\mathbb{N}\); a laundry capacity \(\kappa\in\mathbb{N}\); wear limits \(\theta(s)\in\mathbb{N}\) and loss probabilities \(d(s)\in[0,1]\); a budget \(b\in\mathbb{N}\); and a threshold \(K\in\mathbb{Q}\).
Question: does there exist (i) an initial purchase multiset \(S(0)\subseteq \mathbb{S}\) such that \(\sum_{s\in S(0)} p(s)\le b\), and (ii) a sequence of \(T\) daily choices of two distinct socks \((s_t^1,s_t^2)\in S(t)\times S(t)\) feasible under the dynamics (wear, laundry, disappearance), such that
\[
\sum_{t=1}^{T} \xi(s_t^1,s_t^2)\ \ge\ K\ ?
\]
\end{definition}

A certificate consists of the chosen purchase multiset \(S(0)\) and the sequence of daily pairs \(\{(s_t^1,s_t^2)\}_{t=1}^{T}\). Given the certificate, we can simulate the dynamics for \(T\) steps and verify feasibility and the total compatibility in time polynomial in \(|\mathbb{S}|+T\). Hence, \textsc{Sock-Plan} is in NP.

To this end, we reduce from the \textsc{0--1 Knapsack} decision problem \cite{GareyJohnson1979}. An instance is given by items \(I=\{1,\dots,n\}\), where each item \(i\) has weight \(w_i\in\mathbb{N}\) and value \(v_i\in\mathbb{N}\), a capacity \(W\in\mathbb{N}\), and a target value \(V\in\mathbb{N}\). The question is whether there exists a subset \(I'\subseteq I\) such that \(\sum_{i\in I'} w_i \le W\) and \(\sum_{i\in I'} v_i \ge V\). Formally, given a knapsack instance, we construct a \textsc{Sock-Plan} instance as follows.
\begin{itemize}
    \item \textbf{Item socks.} For each item \(i\in I\), create two socks \(s_i^L\) and \(s_i^R\). Set prices
    \[
    p(s_i^L)=p(s_i^R)=w_i.
    \]
    Set wear limits \(\theta(s)=1\) for all socks and disappearance \(d(s)=0\).
    
    \item \textbf{Filler socks.} Create \(2n\) additional filler socks \(f_1,\dots,f_{2n}\) with price \(p(f_j)=1\), wear limit \(\theta(f_j)=1\), and \(d(f_j)=0\).
    
    \item \textbf{Horizon and laundry.} Set \(T=n\). Set \(\kappa := 2T+1\). Since at most \(2T\) socks can be worn over \(T\) days, the laundry process never triggers, and any worn sock is effectively removed from availability for the remainder of the horizon (it remains in \(L\)).
    
    \item \textbf{Compatibility.} Let \(V_{\text{tot}}:=\sum_{i=1}^{n} v_i\). Define compatibility \(\xi\) by
    \[
    \xi(s_i^L,s_i^R)=\xi(s_i^R,s_i^L)=\frac{v_i}{V_{\text{tot}}},\qquad \forall i\in I,
    \]
    and \(\xi(x,y)=0\) for every other unordered pair \(\{x,y\}\) (including any pair containing a filler sock).
    Equivalently, define \(\eta(x,y):=1-\xi(x,y)\). Note that \(\eta\in[0,1]\) and (importantly) \(\eta\) is a metric: all non-identical pairs have distance \(1\) except \((s_i^L,s_i^R)\), whose distance is in \([0,1)\); thus any triangle contains at most one ``short'' edge, and the triangle inequality holds.
    
    \item \textbf{Budget and target.} Set the sock budget to
    \[
    b := 2W + 2n,
    \]
    which is enough to always include all filler socks (cost \(2n\)) and then spend at most \(2W\) on item socks. Set the decision threshold to
    \[
    K := \frac{V}{V_{\text{tot}}}.
    \]
\end{itemize}

We show that the knapsack instance is a \textsc{yes}-instance if and only if the constructed \textsc{Sock-Plan} instance is a \textsc{yes}-instance.

\begin{itemize}
    \item (\(\Rightarrow\)) Suppose there exists \(I'\subseteq I\) with \(\sum_{i\in I'} w_i\le W\) and \(\sum_{i\in I'} v_i\ge V\).
    Purchase all filler socks and, for each \(i\in I'\), purchase \(s_i^L\) and \(s_i^R\).
    The total cost is \(2n + \sum_{i\in I'} 2w_i \le 2n + 2W = b\).
    Over the \(T=n\) days, schedule the pairs \((s_i^L,s_i^R)\) for each \(i\in I'\) on distinct days, and use arbitrary filler-filler pairs on the remaining days.
    Since \(\theta=1\) and laundry never returns socks (\(\kappa>2T\)), each sock is used at most once and the schedule is feasible.
    The achieved total compatibility is
    \[
    \sum_{t=1}^{T}\xi(s_t^1,s_t^2)=\sum_{i\in I'} \frac{v_i}{V_{\text{tot}}} \ge \frac{V}{V_{\text{tot}}}=K,
    \]
    hence the \textsc{Sock-Plan} instance is a \textsc{yes}-instance.

    \item (\(\Leftarrow\)) Suppose the constructed \textsc{Sock-Plan} instance has a feasible solution with total compatibility at least \(K\).
    Since all compatibilities are \(0\) except between the matched item socks \((s_i^L,s_i^R)\), any day contributing positive compatibility must pair \(s_i^L\) with \(s_i^R\) for some \(i\). Let \(I'\) be the set of items whose socks are paired in this way in the schedule.
    The total compatibility is then \(\sum_{i\in I'} \frac{v_i}{V_{\text{tot}}}\ge K=\frac{V}{V_{\text{tot}}}\), hence \(\sum_{i\in I'} v_i\ge V\).
    Moreover, the schedule requires feasibility over \(T\) days; the filler socks ensure feasibility regardless of \(|I'|\), but the budget constraint implies the item socks purchased satisfy
    \[
    \sum_{i\in I'} 2w_i \le b-2n = 2W
    \quad\Rightarrow\quad
    \sum_{i\in I'} w_i \le W.
    \]
    Therefore \(I'\) is a feasible knapsack solution meeting the value target, so the knapsack instance is a \textsc{yes}-instance.
\end{itemize}

The reduction is polynomial in \(n\), and \textsc{Sock-Plan} is in NP and NP-hard; therefore, it is NP-complete. As a consequence, the corresponding optimization version of the sock buying-and-matching problem (maximizing total compatibility under budget and dynamics) is NP-hard, even in the restricted setting \(d(s)=0\), \(\theta(s)=1\), \(\kappa>2T\), and with a metric mismatch function \(\eta\).

\subsubsection*{It is even hard to approximate: proving approximation constraints}
We show that the optimization version of our sock buying-and-matching model is not only NP-hard, but is also \emph{hard to approximate} within a constant factor better than \(1-\tfrac{1}{e}\), even under a highly restricted special case. The proof is by an approximation-preserving reduction from the \say{Budgeted Maximum Coverage} (BMC) problem.

An instance of \textsc{BMC} consists of a universe of elements \(U\), where each element \(u\in U\) has a nonnegative weight \(w(u)\), a collection of sets \(\mathcal{C}=\{C_1,\dots,C_m\}\) with costs \(c_i\in\mathbb{N}\), and a budget \(B\in\mathbb{N}\). The goal is to pick a subcollection \(\mathcal{C}'\subseteq \mathcal{C}\) of total cost at most \(B\) that maximizes the total weight of covered elements:
\[
\max_{\mathcal{C}'\subseteq \mathcal{C}} \ \sum_{u\in \cup_{C\in \mathcal{C}'} C} w(u)
\quad\text{s.t.}\quad \sum_{C_i\in \mathcal{C}'} c_i \le B.
\]
It is known that \textsc{BMC} admits a \((1-\tfrac{1}{e})\)-approximation algorithm and that this factor is essentially tight: improving it by any constant \(\varepsilon>0\) is impossible under standard complexity assumptions \cite{KhullerMossNaor1999}.

To this end, we will use a restricted sock subproblem. Let us consider the following restriction of our model: we only optimize the initial purchase \(S(0)\) under a budget, and the objective is purely a weighted \say{diversity} term. Formally, let each sock \(s\) carry a set of symbolic appearance-features \(F(s)\subseteq U\) (these can be represented as a subset-encoding within \(\overrightarrow{u}\)). Define the weighted diversity functional
\begin{equation}
D_w(S) \ :=\ \sum_{u \in \cup_{s\in S} F(s)} w(u).
\label{eq:weighted_diversity}
\end{equation}
Set the horizon \(T:=1\) and enforce that the day-1 pairing contributes zero utility (e.g., set \(\rho=0\), or define \(\xi\equiv 0\) and \(\chi=0\)), and ignore ecological conversion by setting \(\lambda=0\). Evaluate diversity only at purchase time (i.e., add \(\delta \cdot D_w(S(0))\) once with \(\delta=1\)). The resulting optimization reduces to
\begin{equation}
\max_{S(0)\subseteq \mathbb{S}} D_w(S(0))
\quad \text{s.t.}\quad \sum_{s\in S(0)} p(s) \le b.
\label{eq:sock_design}
\end{equation}
We call this restricted problem \textsc{Sock-Design}.

To this end, given a \textsc{BMC} instance \((U,w,\mathcal{C},c,B)\), construct a \textsc{Sock-Design} instance as follows:
\begin{itemize}
    \item For each set \(C_i\in \mathcal{C}\), create a sock \(s_i\) with price \(p(s_i):=c_i\) and feature-set \(F(s_i):=C_i\).
    \item Set the sock budget \(b:=B\).
    \item Use the weighted diversity objective \(D_w\) from Eq.~(\ref{eq:weighted_diversity}).
\end{itemize}
Then, for any chosen sock subset \(S(0)\), the objective value in Eq.~(\ref{eq:sock_design}) equals the total weight of elements covered by the corresponding chosen sets in \textsc{BMC}, and the budget constraints match exactly. Hence, the transformation is polynomial time and approximation preserving.

Thus, suppose there were a polynomial-time \(\alpha\)-approximation algorithm for the full sock optimization problem (and therefore for its restriction \textsc{Sock-Design}) with \(\alpha > 1-\tfrac{1}{e}+\varepsilon\) for some \(\varepsilon>0\). By the reduction above, this would imply a polynomial-time \(\alpha\)-approximation algorithm for \textsc{BMC}, contradicting the known inapproximability threshold for \textsc{BMC} \cite{KhullerMossNaor1999}. Therefore, for every \(\varepsilon>0\), achieving an approximation ratio better than \(1-\tfrac{1}{e}+\varepsilon\) for our sock optimization problem is impossible (under the same standard complexity assumptions as for \textsc{BMC}), even when restricted to instances in which the dynamics and daily pairing are neutralized and only the initial sock acquisition is optimized via the weighted diversity objective.

\subsection*{Simulation initialization and reference configuration}
\label{sec:app_sim_init}
Each simulation run is initialized by (i) sampling a sock catalogue of designs from Amazon, (ii) purchasing an initial inventory under a fixed budget, and (iii) initializing the dynamic state variables.

A catalogue comprises \(n_{\text{designs}}=1248\) sock designs collected from Amazon by searching for \say{socks}. Each design \(j\in\{1,\dots,n_{\text{designs}}\}\) is assigned a discrete appearance feature vector
\[
\overrightarrow{u}_j \in \mathcal{U} = \{0,\dots,m_1-1\}\times\cdots\times\{0,\dots,m_k-1\},
\]
where \((m_1,\dots,m_k)=\texttt{feature\_sizes}\) and \(k=|\texttt{feature\_sizes}|\). In \(\Theta_{\text{ref}}\), \(k=3\), corresponding to (color, pattern, length) with \(m=(32,13,3)\). Compatibility between designs is computed via normalized Hamming dissimilarity:
\[
\eta(j,\ell)=\frac{1}{k}\sum_{r=1}^{k}\mathbb{I}\left[u_{j,r}\neq u_{\ell,r}\right],\qquad
\xi(j,\ell)=1-\eta(j,\ell),
\]
where \(\eta\in[0,1]\) and \(\xi\in[0,1]\). Thus, designs that differ in fewer appearance components are more compatible. Each design receives an integer price computed for a single sock (in cases where sets of socks are listed together).

At \(t=0\), the simulator purchases socks in pairs until the monetary budget is exhausted. The diversity preference parameter \(\delta\in[0,1]\) controls how concentrated the purchases are across designs. Concretely, the simulator samples an \say{active} subset of designs of size
\[
k_{\text{active}} = 1 + \left\lfloor \delta\cdot (n_{\text{designs}}-1) \right\rceil,
\]
and repeatedly buys pairs from this active set (subject to affordability). Therefore, \(\delta=0\) produces highly uniform inventories (nearly all socks from one design), while \(\delta\approx 1\) approaches near-uniform sampling across the whole catalogue.

After purchases, all socks are initialized with \(\tau=0\), wear limit \(\theta\), and wash-disappearance probability \(d\). The laundry buffer is initialized empty, \(L(0)=\emptyset\), and the available inventory is \(S(0)\). If replenishment is enabled, purchases may occur during the run when inventory becomes too small to form a pair, subject to remaining budget.

Table~\ref{tab:theta_ref} lists the reference configuration used to generate the draft results in Section~\ref{sec:results}. The setting corresponds to a one-year horizon with weekly-ish laundry batching, low per-wash disappearance probability, moderate public exposure, and a medium-diversity initial buying preference.

\begin{table}[!ht]
\centering
\begin{tabular}{lll}
\hline \hline
Category & Parameter & Value / interpretation \\
\hline \hline
Time & \(T\) & 365 days (horizon) \\
Laundry & \(\kappa\) & 14 socks (batch size threshold) \\
Budget & \(b\) & 200\$ (total spend limit) \\
Wear-out & \(\theta\) & 50 wears before unwearable \\
Disappearance & \(d\) & 0.02 per wash event per sock \\
Exposure & \(\rho\) & 0.50 probability a day is \say{public} \\
Sensitivity & \(\chi\) & 1.25 mismatch sensitivity \\
Penalty shape & \(\gamma\) & 1.02 in \(g(\eta)=\eta^{\gamma}\) \\
Eco proxy & \(\alpha\) & 1.0 in \(e=\alpha p\) \\
Buying diversity & \(\delta\) & 0.50 (medium concentration) \\
\hline\hline
\end{tabular}
\caption{Simulator initialization parameters.}
\label{tab:theta_ref}
\end{table}
\subsection*{Sensitivity to the ecological-cost conversion rate}
\label{app:alpha_sensitivity}
Since the reference simulator uses the proportional ecological proxy \(e(s)=\alpha p(s)\), we tested whether the conclusions depend on the selected conversion rate. We varied \(\alpha\) from \(0.75\) to \(1.25\) in increments of \(0.05\) and recomputed the reference-regime outcomes. Table~\ref{tab:alpha_sensitivity} reports the ecological-cost proxy together with two operational metrics: infeasible days and stranded wear-capacity. As expected, \(C_{\text{eco}}\) varies with \(\alpha\). However, infeasible days and stranded wear-capacity remain unchanged because \(\alpha\) only rescales the ecological-cost proxy and does not enter the pairing policy, laundry-loss process, or wear dynamics. This indicates that the main operational conclusions about mismatch tolerance are not sensitive to the selected ecological conversion rate.

\begin{table}[!ht]
\centering
\begin{tabular}{cccc}
\hline \hline
\(\alpha\) 
& \(C_{\text{eco}}\) 
& Infeasible days 
& \(C_{\text{strand}}\) \\
\hline \hline
0.75 & \(149.79 \pm 0.43\) & \(23.23 \pm 11.40\) & \(1051.46 \pm 128.08\) \\
0.80 & \(159.77 \pm 0.46\) & \(23.23 \pm 11.40\) & \(1051.46 \pm 128.08\) \\
0.85 & \(169.76 \pm 0.48\) & \(23.23 \pm 11.40\) & \(1051.46 \pm 128.08\) \\
0.90 & \(179.74 \pm 0.51\) & \(23.23 \pm 11.40\) & \(1051.46 \pm 128.08\) \\
0.95 & \(189.73 \pm 0.54\) & \(23.23 \pm 11.40\) & \(1051.46 \pm 128.08\) \\
1.00 & \(199.72 \pm 0.57\) & \(23.23 \pm 11.40\) & \(1051.46 \pm 128.08\) \\
1.05 & \(209.70 \pm 0.60\) & \(23.23 \pm 11.40\) & \(1051.46 \pm 128.08\) \\
1.10 & \(219.69 \pm 0.63\) & \(23.23 \pm 11.40\) & \(1051.46 \pm 128.08\) \\
1.15 & \(229.67 \pm 0.66\) & \(23.23 \pm 11.40\) & \(1051.46 \pm 128.08\) \\
1.20 & \(239.66 \pm 0.68\) & \(23.23 \pm 11.40\) & \(1051.46 \pm 128.08\) \\
1.25 & \(249.64 \pm 0.71\) & \(23.23 \pm 11.40\) & \(1051.46 \pm 128.08\) \\
\hline \hline
\end{tabular}
\caption{Sensitivity to the ecological-cost conversion rate \(\alpha\). Values are reported as mean \(\pm\) standard deviation across the reference-regime policy means. The ecological-cost proxy \(C_{\text{eco}}\) changes with \(\alpha\), while infeasible days and stranded wear-capacity remain stable because \(\alpha\) does not affect policy decisions or sock dynamics.}
\label{tab:alpha_sensitivity}
\end{table}

\end{document}